\documentclass[11pt]{amsart}
\usepackage{latexsym,amssymb,amsmath,amscd,amsthm}
\numberwithin{equation}{section}
\theoremstyle{remark}

\newcommand{\bq}{\begin{equation}}
\newcommand{\bea}{\begin{array}}
\newcommand{\eea}{\end{array}}

\newcommand{\ga}{\alpha}
\newcommand{\gep}{\epsilon}
\newcommand{\gD}{\Delta}
\newcommand{\gl}{\lambda}
\newcommand{\gL}{\Lambda}
\newcommand{\gb}{\beta}

\newcommand{\mf}{\mathfrak}

\newcommand{\go}{\omega}
\newcommand{\gO}{\Omega}
\newcommand{\gG}{\Gamma}

\newcommand{\gs}{\sigma}

\newcommand{\gag}{\gamma}
\newcommand{\gd}{\delta}
\newcommand{\pp}{\partial}

\newcommand{\tl}{\tilde}
\newcommand{\na}{\nabla}
\newcommand{\gk}{\kappa}

\newcommand{\bs}{\blacksquare}

\newcommand{\gT}{\Theta}

\newcommand{\bgs}{\bigstar}

\newcommand{{\DDD}}{D\!\!\!\!\!\!-}

\newcommand{\bx}{\Box}

\title{REMARKS ON WEYL GEOMETRY AND QUANTUM MECHANICS}

\author{Robert Carroll\\University of Illinois, Urbana, IL 61801}

\date{January, 2008\thanks{email: rcarroll@math.uiuc.edu}}

\begin{document}

\bibliographystyle{plain}

\begin{abstract} 
A short survey of some material related to conformal general relativity (CGR),
integrable Weyl geometry, and Dirac-Weyl (DW) theory is given which suggests that
CGR is essentially equivalent to DW with quantum mass corresponding to 
conformal mass; furthermore various actions can be reformulated in terms of the quantum potential.  
\end{abstract}

\maketitle
\tableofcontents


\section{INTRODUCTION}
\renewcommand{\theequation}{1.\arabic{equation}}
\setcounter{equation}{0}

This is primarily a survey and review article, the main point of which is to make
precise an observation of Bonal, Quiros, and Cardenas in \cite{bol} relating
conformal mass to quantum mass.  We acknowledge with thanks the valuable
comments of a reviewer, some of which are partially included here, and some
typos have been corrected.  A revised and
shortened version is in preparation.
\\[3mm]\indent
Thus we will gather together some formulas involving Weyl geometry and
Weyl-Dirac theory in various ways and show connections to the
Schr\"odinger equation (SE) and Klein-Gordon equation (KG).  
Many aspects of this have been sketched in \cite{c1,c2,c3,c5} 
(based in part on \cite{au,av,aw,c16,c17,c18,i1,i2,s35,ti,w1}) and related it to Ricci
flow in \cite{c4} and we now want to deepen our understanding and
rephrase some material with some expansion.
The presentations in \cite{dt,schz,sczh} are very extensive so we will begin
with Israelit \cite{i1,i2} to get started (cf. also \cite{o1} for semi-Riemannian 
geometry).
We recall first the standard transformations of contravariant and covariant
vectors via $x\to \bar{x}$ in the form
\bq\label{1.1}
T^{\nu}\to\bar{T}^{\nu}=T^{\gs}\frac{\pp\bar{x}^{\nu}}{\pp x^{\gs}};\,\,T_{\mu}\to
\bar{T}_{\mu}=T_{\gs}\frac{\pp x^{\gs}}{\pp\bar{x}^{\mu}}
\end{equation}
For parallel transport of a vector along a curve from $x^{\nu}\to x^{\nu}+dx^{\nu}$
one sets
\bq\label{1.2}
dT^{\mu}=-T^{\gs}\gG^{\mu}_{\gs\nu}dx^{\nu};\,\,\gG^{\gl}_{\mu\nu}=
\frac{1}{2}g^{\gl\gs}[\pp_{\nu}g_{\mu\gs}+\pp_{\mu}g_{\nu\gs}-\pp_{\gs}
g_{\mu\nu}]
\end{equation}
where $\gG^{\gl}_{\mu\nu}$ is the standard Christoffel symbol.
Some calculation then shows that $dT=0$ under parallel transport and
one obtains also $({\bf 1A})\,\,dT_{\mu}=T_{\gs}\gG^{\gs}_{\mu\nu}dx^{\nu}$.
In Riemannian geometry one defines also covariant derivatives via
\bq\label{1.3}
\na_{\nu}T^{\mu}=\pp_{\nu}T^{\mu}+T^{\gs}\gG^{\mu}_{\gs\nu};\,\,
\na_{\nu}T_{\mu}=\pp_{\nu}T_{\mu}-T_{\gs}\gG^{\gs}_{\mu\nu}
\end{equation}
From the above one determines easily that $({\bf 1B})\,\,\na_{\gl}g_{\mu\nu}=0
=\na_{\gl} g^{\mu\nu}$ and if the vector $T^{\mu}$ is parallel transported 
around an infinitesimal closed parallelogram it follows that $(\bullet)\,\,
\gD T^{\gl}=T^{\gs}R^{\gl}_{\gs\mu\nu}dx^{\mu}dx^{\nu}$ where the Riemann-
Christoffel tensor is
\bq\label{1.4}
R^{\gl}_{\gs\mu\nu}=-\pp_{\nu}\gG^{\gl}_{\gs\mu}+\pp_{\mu}\gG^{\gl}_{\gs\nu}
-\gG^{\ga}_{\gs\mu}\gG^{\gl}_{\ga\nu}+\gG^{\ga}_{\gs\nu}\gG^{\gl}_{\ga\mu}
\end{equation}
On the other hand from $dT=0$ the length of the vector $\gD T=0$ and one 
finds
\bq\label{1.5}
\na_{\nu}\na_{\mu}T_{\gl}-\na_{\mu}\na_{\nu}T_{\gl}=T_{\gs}R^{\gs}_{\gl\mu\nu}
\end{equation}
Contracting $R^{\gs}_{\mu\nu\gl}$ gives $R_{\mu\nu}=R^{\gl}_{\mu\nu\gl}$ and the curvature scalar $R=g^{\mu\nu}R_{\mu\nu}$.
\\[3mm]\indent
For Weyl geometry one has a metric tensor $g_{\mu\nu}=g_{\nu\mu}$ and a
length connection vector $w_{\mu}$ while both the direction and the length of a vector change under parallel transport.  Thus if a vector is displaced 
by $dx^{\nu}$ one has $({\bf 1C})\,\,dT^{\mu}=-T^{\gs}\hat{\gG}^{\mu}_{\gs\nu}
dx^{\nu}$ and its length is changed via $({\bf 1D})\,\,dT=Tw_{\nu}dx^{\nu}$.
One can also write $({\bf 1E})\,\,d(T^2)=2T^2w_{\nu}dx^{\nu}$ and for this to
agree with ({\bf 1C}) and $T^2=g_{\mu\nu}T^{\mu}T^{\nu}$ 
for arbitrary $T^{\mu}$ and $dx^{\nu}$ requires
\bq\label{1.6}
g_{\mu\gs}\hat{\gG}^{\gs}_{\nu\gl}+g_{\nu\gs}\hat{\gG}^{\gs}_{\mu\gl}=\pp_{\gl}g_{\mu\nu}
-2g_{\mu\nu}w_{\gl}
\end{equation}
Following Weyl one assumes that the connection is symmetric (i.e.
$\hat{\gG}^{\gl}_{\mu\nu}=\hat{\gG}^{\gl}_{\nu\mu}$) leading to
\bq\label{1.7}
\hat{\gG}^{\gl}_{\mu\nu}=\gG^{\gl}_{\mu\nu}+g_{\mu\nu}w^{\gl}-\gd_{\nu}^{\gl}w_{\mu}-\gd^{\gl}_{\mu}w_{\nu}
\end{equation}
Using this Weyl connection $\hat{\gG}^{\gl}_{\mu\nu}$ one forms a 
covariant Weyl derivative via
\bq\label{1.8}
\hat{\na}_{\nu}T^{\mu}=\pp_{\nu}T^{\mu}+T^{\gs}\hat{\gG}^{\mu}_{\gs\nu};\,\,
\hat{\na}_{\nu}T_{\mu}=\pp_{\nu}T_{\mu}-T_{\gs}\hat{\gG}^{\gs}_{\mu\nu}
\end{equation}
One can also write then
\bq\label{1.9}
\hat{\na}_{\nu}T^{\mu}=\na_{\nu}T^{\mu}+T^{\gs}[g_{\gs\nu}w^{\mu}-
\gd_{\nu}^{\mu}w_{\gs}-\gd_{\gs}^{\mu}w_{\nu}]
\end{equation}
This all generalizes to tensors in an obvious manner and in particular
\bq\label{1.10}
\hat{\na}_{\gl}g_{\mu\nu}=\na_{\gl}g_{\mu\nu}-g_{\gs\nu}\hat{\gG}^{\gs}_{\mu\gl}
-g_{\mu\gs}\hat{\gG}^{\gs}_{\nu\gl}=2g_{\mu\nu}w_{\gl}
\end{equation}
Similarly one has $({\bf 1F})\,\,\na^W_{\gl}g^{\mu\nu}=-2g^{\mu\nu}w_{\gl}$.
\\[3mm]\indent
In particular consider a spacetime with a symmetric metric 
tensor $g_{\mu\nu}$ and an asymmetric connection $\tl{\gG}^{\gl}_{\mu\nu}$
defining a parallel displacement (cf. ({\bf 1C}))
$({\bf 1G})\,\,dT^{\mu}=-T^{\gs}\tl{\gG}^{\mu}_{\gs\nu}dx^{\nu}$.  The connection
can be split into three parts, the Christoffel symbol, the contorsion tensor, and the 
nonmetricity, namely 
\bq\label{1.11}
\tl{\gG}^{\gl}_{\mu\nu}=\gG^{\gl}_{\mu\nu}+C^{\gl}_{\mu\nu}+\frac{1}{2}g^{\gl\gs}
[Q_{\nu\mu\gs}+Q_{\mu\nu\gs}-Q_{\gl\mu\nu}]
\end{equation}
The contorsion tensor $C^{\gl}_{\mu\nu}$ may be expressed in terms of the
torsion tensor $\tl{\gG}^{\gl}_{[\mu\nu]}=(1/2)[\tl{\gG}^{\gl}_{\mu\nu}-
\tl{\gG}^{\gl}_{\nu\mu}]$ via
$({\bf 1H})\,\,C^{\gl}_{\mu\nu}=\tl{\gG}^{\gl}_{[\mu\nu]}+ g^{\gl\gs}
g_{\mu\gk}\tl{\gG}^{\gk}_{[\nu\gs]}+g^{\gl\gs}g_{\nu\gk}\tl{\gG}^{\gk}_{[\mu\gs]}$.  The
nonmetricity can in turn be represented as minus the covariant derivative of the
metric tensor with respect to $\tl{\gG}^{\gl}_{\mu\nu}$, namely
\bq\label{1.12}
Q_{\gl\mu\nu}=-\pp_{\gl}g_{\mu\nu}+g_{\gs\nu}\tl{\gG}^{\gs}_{\mu\gl}+
g_{\gs\mu}\tl{\gG}^{\gs}_{\nu\gl}
\end{equation}
Comparing with the Weyl connection \eqref{1.7} one can say that the Weyl
geometry is torsionless with nonmetricity $({\bf 1I})\,\,\hat{Q}_{\gl\mu\nu}=
-2g_{\mu\nu}w_{\gl}$ (in Riemannian geometry both torsion and nommetricit
vanish).
Further in the Weyl geometry for parallel transport of a vector $T^{\mu}$ around an infinitesimal closed parallelogram one has
$({\bf 1J})\,\,\gD T^{\gl}=T^{\gs}K^{\gl}_{\gs\mu\nu}dx^{\mu}\gd x^{\nu}$
and using ({\bf 1D}) there results for the total length change $({\bf 1K})\,\,
\gD T=TW_{\mu\nu}dx^{\mu}\gd x^{\nu}$.  In ({\bf 1J}) $K^{\gl}_{\gs\mu\nu}$
is a curvature tensor similar to the Riemannian one but with $\gG^{\gl}_{\mu\nu}$
replaced by the Weyl connection $\hat{\gG}^{\gl}_{\mu\nu}$.  In ({\bf 1K})
one has the Weyl length curvature tensor $({\bf 1M})\,\,W_{\mu\nu}=
\na_{\nu}w_{\mu}-\na_{\mu}w_{\nu}=\pp_{\nu}w_{\mu}-\pp_{\mu}w_{\nu}$
and one sees that in Weyl geometry $\gD T\ne 0$ unless $W_{\mu\nu}=0$.
\\[3mm]\indent
Transporting a vector of length T around a closed loop leads to $T_{new}=
T+\int_STW_{\mu\nu}dS^{\mu\nu}$ where S is the area confined by the loop
and $dS^{\mu\nu}$ a suitable area element.  If one assumes that under Weyl 
gauge transformations (WGT) the components $T^{\mu}$ remain unchanged
the length changes via $T\to\tl{T}=exp(\gl)T$ where $\gl(x^{\nu})$ is a
differentiable function of the coordinates.  One has then
\bq\label{1.13}
g_{\mu\nu}\to \tl{g}_{\mu\nu}=e^{2\gl}g_{\mu\nu};\,\,g^{\mu\nu}\to\tl{g}^{\mu\nu}
= e^{-2\gl}g^{\mu\nu}
\end{equation}
In order to have $dT=Tw_{\nu}dx^{\nu}$ for transformed quantities one has to
take a WGT law $({\bf 1N})\,\,w_{\nu}\to\tl{w}_{\nu}=w_{\nu}+\pp_{\nu}\gl$.
Further a quantity is said to be gauge covariant if under a WGT it is transformed
via $({\bf 1O})\,\,\psi\to\tl{\psi}=exp(n\gl)\psi$ and $n$ is called the Weyl power of
$\psi$; one writes $\Pi(\psi)=n$.  Thus $\Pi(g_{\mu\nu})=2,\,\,\Pi(g^{\mu\nu})
=-2,\,\,\Pi(T)=1,\,\,\Pi(\sqrt{-g})=4,$ etc.  If $\Pi$ is zero one has a gauge invariant
quantity and making use of \eqref{1.13} and ({\bf 1N}) it follows that 
$({\bf 1P})\,\,\Pi(\hat{\gG}^{\gl}_{\mu\nu})=0$; quantities are called
in-invariant if they are both coordinate and gauge invariant.
\\[3mm]\indent
{\bf REMARK 1.1.}
The Dirac modification of Weyl theory involves a scalar function $\gb(x^{\nu})$
satisfying $({\bf 1Q})\,\,\gb\to\tl{\gb}=exp(-\gl)\gb$ (i.e. $\Pi(\gb)=-1$).  Since
this provides a 1-1 relation between $\gl$ and $\gb$ one calls $\gb$ the Dirac
gauge function (also Dirac field).  Given that $\gl$ is intrinsic to Weyl spacetime
one can say that $\gb$ is also a part of the geometric structure (cf. also
\cite{d1,i3,r1}).$\hfill\bs$
\\[3mm]\indent
Now some curvature tensors are developed starting with \eqref{1.4}-\eqref{1.5}.
Contracting $\hat{R}^{\gl}_{\gs\mu\nu}$ one obtains the Ricci tensor
\bq\label{1.14}
\hat{R}_{\mu\nu}=\hat{R}^{\gl}_{\mu\nu\gl}=-\pp_{\gl}\gG^{\gl}_{\mu\nu}+
\pp_{\nu}\gG^{\gl}_{\mu\gl}-\gG^{\ga}_{\mu\nu}\gG^{\gl}_{\ga\gl}+
\gG^{\ga}_{\mu\gl}\gG^{\gl}_{\ga\nu}
\end{equation}
This leads to the Ricci curvature $R=g^{\mu\nu}R_{\mu\nu}$ and the Einstein
tensor  $({\bf 1R})\,\,G^{\nu}_{\mu}=R^{\nu}_{\mu}-(1/2)\gd_{\mu}^{\nu}R$
which satisfies the contracted Bianchi 
identity $({\bf 1S})\,\,\na_{\nu}G^{\nu}_{\mu}=0$.  For the Weyl counterpart
we have following \eqref{1.3} $({\bf 1T})\,\,\hat{\na}_{\nu}\hat{\na}_{\mu}
T_{\gl}-\hat{\na}_{\mu}\hat{\na}_{\nu}T_{\gl}=T_{\gs}K^{\gs}_{\gl\mu\nu}$ with
Weylian curvature tensor $K^{\gs}_{\gl\mu\nu}$ given by
\bq\label{1.15}
K^{\gl}_{\gs\mu\nu}=-\pp_{\nu}\hat{\gG}^{\gl}_{\gs\mu}+\pp_{\mu}
\hat{\gG}^{\gl}_{\gs\nu}-\hat{\gG}^{\ga}_{\gs\mu}\hat{\gG}^{\gl}_{\ga\nu}
+\hat{\gG}^{\ga}_{\gs\nu}\hat{\gG}^{\gl}_{\ga\mu}
\end{equation}
Contracting this and using \eqref{1.7} yields
\bq\label{1.16}
K_{\mu\nu}=K^{\gl}_{\mu\nu\gl}=
\end{equation}
$$R_{\mu\nu}-g_{\mu\nu}\na_{\gl}w^{\gl}+\na_{\mu}w_{\nu}-3\na_{\nu}w_{\mu}+2g_{\mu\nu}w^{\gl}w_{\gl}-2w_{\mu}w_{\nu}$$
Consequently 
\bq\label{1.17}
K_{\mu\nu}-K_{\nu\mu}=4(\na_{\mu}w_{\nu}-\na_{\nu}w_{\mu})=-4W_{\mu\nu}
\end{equation}
Contracting in \eqref{1.15} one obtains now
$({\bf 1U})\,\,K^{\gl}_{\gl\mu\nu}=-4W_{\mu\nu}$ and from \eqref{1.16} there is
a scalar $({\bf 1V})\,\,K=g^{\mu\nu}K^{\gl}_{\mu\nu\gl}=R-6\na_{\gl}w^{\gl}+
6w^{\gl}w_{\gl}$.  One forms also $({\bf 1W})\,\,W^2=W_{\gl\gs}W^{\gl\gs}$ and considers a Weyl space with torsion.  In this case the connection can be written
in the form
\bq\label{1.18}
\bar{\gG}^{\gl}_{\mu\nu}=\gG^{\gl}_{\mu\nu}+g_{\mu\nu}w^{\gl}-\gd_{\nu}^{\gl}
w_{\mu}-\gd_{\mu}^{\gl}w_{\nu}+C^{\gl}_{\mu\nu}
\end{equation}
and one defines a covariant derivative $({\bf 1X})\,\,\bar{\na}_{\nu}T_{\gl}=
\pp_{\nu}T_{\gl}-T_{\gs}\bar{\gG}^{\gs}_{\gl\nu}$ with
\bq\label{1.19}
\bar{\na}_{\nu}\bar{\na}_{\mu}T_{\gl}-\bar{\na}_{\mu}\bar{\na}_{\nu}T_{\gl}
=T_{\gs}\bar{K}^{\gs}_{\gl\mu\nu}-2\bar{\na}_{\gs}T_{\gl}\bar{\gG}^{\gs}_{[\mu\nu]}
\end{equation}
\indent
In summary one notes that in Riemannian geometry under parallel transport
the direction of a vector is changed but not its length.  However in Weyl 
geometry both the length and direction are changed.  In theories based on Weyl
geometry the equations and laws have to be covariant with respect to both
coordinate transformations (CT) and Weyl gauge transformations (WGT).
Both Weyl connections $\hat{\gG}$ and $\bar{\gG}$ are gauge invariant so the
corresponding contractions and curvatures are covariant with respect to CT
and WGT.

\section{THE DIRAC WEYL APPROACH}
\renewcommand{\theequation}{2.\arabic{equation}}
\setcounter{equation}{0}

First, following Israelit \cite{i1}, we add a few more remarks about Weyl geometry.
Look first at the Weyl geometry without torsion 
involving $\hat{\gG}^{\gl}_{\mu\nu}$ with nonmetricity $Q_{\gl\mu\nu}$ as in
\eqref{1.12}.  There are two other possible choices of connection of interest here.
\begin{enumerate}
\item
For ${}^2\hat{\gG}^{\gl}_{\mu\nu}=\gG^{\gl}_{\mu\nu}+g_{\mu\nu}w^{\gl}
+\gd^{\gl}_{\mu}w_{\nu}-\gd^{\gl}_{\nu}w_{\mu}$ with ({\bf 1C}) 
(cf. \eqref{1.7}) there results
$d_2T=-Tw_{\nu}dx^{\nu}$ and ${}^2Q_{\gl\mu\nu}=2g_{\mu\nu}w_{\gl}$
(cf. ({\bf 1I})) leading to a torsion tensor ${}^2\hat{\gG}^{\gl}_{[\mu\nu]}=\gd^{\gl}_{\mu}
w_{\nu}-\gd^{\gl}_{\nu}w_{\mu}$
\item
Another connection is possible with the form ${}^3\hat{\gG}^{\gl}_{\mu\nu}=
\gG^{\gl}_{\mu\nu}+(1/2){}^2C^{\gl}_{\mu\nu}$ where ${}^2C^{\gl}_{\mu\nu}=
2g_{\mu\nu}w^{\gl}-2\gd^{\gl}_{\nu}w_{\mu}$ is the cotorsion tensor for 
${}^2\hat{\gG}^{\gl}_{\mu\nu}$.  Here one has $d_3T=0=\gD_3T$ and vanishing
nonmetricity, but there is torsion.
\end{enumerate}

\indent
Now the Weyl-Dirac theory was developed in various ways following \cite{d1,i1,i2,r1,s35}
(cf. also \cite{c1,c5} for a survey) and we follow first \cite{i1} here.  Thus one can deal with an action integral (cf. \cite{i1} for details)
\bq\label{2.1}
I=\int [W^{\gl\mu}W_{\gl\mu}-\gb^2R+\gs\gb^2w^{\gl}w_{\gl}+(\gs+6)\pp_{\gl}\gb
\pp^{\gl}\gb+
\end{equation}
$$+2\gs\gb\pp_{\gl}\gb w^{\gl}+2\gL\gb^4+L_M]\sqrt{-g}d^4x$$
($\gL$ is a cosmological constant).
Note here ($L_G=L_{geom}$)
$$(\bgs)\,\,-\gb^2K=-\gb^2R+6\gb^2\na_{\gl}w^{\gl}-6\gb^2w^{\gl}w_{\gl}$$
$$(\bgs\bgs)\,\,L_G=W^{\gl\gs}W_{\gl\gs}-\gb^2R+6\gb^2\na_{\gl}w^{\gl}
-6\gb^2w^{\gl}w_{\gl}+k(\pp_{\gl}\gb+\gb w_{\gl})^2+2\gL\gb^4$$
However $\gb^2\na_{\gl}w^{\gl}=\na_{\gl}(\gb^2w^{\gl})-2\gb\pp_{\gl}\gb w^{\gl}$ and
the first term integrates out, leading to \eqref{2.1}.
Then to avoid Proca terms one sets $\gs=k-6=0$ leading to
\bq\label{2.2}
I=\int[W^{\gl\mu}W_{\gl\mu}-\gb^2R+6\pp_{\gl}\gb\pp^{\gl}\gb +2\gL\gb^4+L_M]
\sqrt{-g}d^4x
\end{equation}
(cf. \cite{i1} for details).
Here $\gb$ is an additional dynamical variable (field) with $\Pi(\gb)=-1$ and
one is interpreting $w_{\mu}$ as an EM field potential with $W_{\gl\mu}$ as
the EM field strength.  The equations of the theory follow from $\gd I=0$ and first
varying $g_{\mu\nu}$ gives
\bq\label{2.3}
G^{\mu\nu}=R^{\mu\nu}-\frac{1}{2}g^{\mu\nu}R=
-\frac{8\pi}{\gb^2}(M^{\mu\nu}+T^{\mu\nu})+
\end{equation}
$$+\frac{2}{\gb}(g^{\mu\nu}\na^{\ga}\na_{\ga}\gb-\na^{\nu}\na^{\mu}\gb)+
\frac{1}{\gb^2}(4\pp^{\mu}\gb \pp^{\nu}\gb-g^{\mu\nu}\pp_{\ga}\gb\pp^{\ga}\gb)
-g^{\mu\nu}\gL\gb^2$$
One has energy-momentum tensors $({\bf 2B})\,\,M^{\mu\nu}=(1/4\pi)
[(1/4)g^{\mu\nu}W^{\gl\gs}W_{\gl\gs}-W^{\mu\gl}W^{\nu}_{\gl}]$ for the EM field and
$({\bf 2C})\,\,T^{\mu\nu}=[(1/8\pi \sqrt{-g})[\gd(\sqrt{-g}L_M)/\gd g_{\mu\nu}]$ for the 
matter.  Varying $w_{\mu}$ gives the Maxwell equation 
$({\bf 2D})\,\,\na_{\nu}W^{\mu\nu}=4\pi J^{\mu}$ where $J^{\mu}=(1/16\pi)(\gd L_M/
\gd w_{\mu})$ and varying in $\gb$ one obtains
\bq\label{2.4}
\gb R+6\na^{\gl}\na_{\gl}\gb-4\gL\gb^3=8\pi B;\,\,B=\frac{1}{16\pi}\frac{\gd L_M}{\gd\gb}
\end{equation}
(B is a "charge" conjugate to $\gb$).  Taking the divergence of ({\bf 2D}) gives the
conservation law for electric charge $({\bf 2E})\,\,\na_{\gl}J^{\gl}=0$ and taking the
divergence of \eqref{2.3} one gets (using the Bianchi identity) the equations of motion
for the matter 
$({\bf 2F})\,\,\na_{\nu}T^{\mu\nu}-T(\pp^{\mu}\gb/\gb)=-W^{\mu\nu}J_{\nu}$.
\\[3mm]\indent
Consider now the matter part of the in-invariant action \eqref{2.2}, namely
$({\bf 2G})\,\,I_M=\int L_M(\sqrt{-g}d^4x$ which depends on $g_{\mu\nu},\,\,w_{\mu},\,\,
\gb,$ and perhaps additional dynamical variables $\psi_{\mu}$.  Since the latter 
are not present in $L_M$ one has $({\bf 2H})\,\,\gd[\sqrt{-g}L_M]/\gd\psi_{\mu}=0$
so 
\bq\label{2.5}
\gd I_M=8\pi\int [T^{\mu\nu}\gd g_{\mu\nu}+2J^{\mu}\gd w_{\mu}+2B\gd\gb]
\sqrt{-g}d^4x
\end{equation}
Now carry out a CT with an arbitrary infinitesimal vector $\eta^{\mu}$, i.e. $x^{\mu}\to
\bar{x}^{\mu}=x^{\mu}+\eta^{\mu}$; this yields
\bq\label{2.6}
\gd g_{\mu\nu}=g_{\gl\nu}\na_{\mu}\eta^{\gl}+g_{\mu\gl}\na_{\nu}\eta^{\gl};\,\,
\gd w_{\mu}=w_{\gl}\na_{\mu}\eta^{\gl}+\na_{\gl}w_{\mu}\eta^{\gl};\,\,\gd \gb
=\pp_{\gl}\gb\,\eta^{\gl}
\end{equation}
Putting this in \eqref{2.5} and integrating by parts gives
\bq\label{2.7}
\gd I_M=16\pi\int [-\na_{\gl}T^{\gl}_{\mu}-w_{\mu}\na_{\gl}J^{\gl}-J^{\gl}W_{\mu\gl}
+B\pp_{\mu}\gb]\eta^{\mu}\sqrt{-g}d^4x
\end{equation}
However since $I_M$ is an in-invariant its variation with respect to CT vanishes so 
from \eqref{2.7} one obtains the conservation law $({\bf 2J})\,\,\na_{\gl}T^{\gl}_{\mu}
+w_{\mu}\na_{\gl}J^{\gl}+J^{\gl}W_{\mu\gl}-B\pp_{\mu}\gb=0$.  Next consider a 
WGT with $\gl(x^{\mu})$ infinitesimal.  Using \eqref{1.13}, ({\bf 1N}), and ({\bf 1Q})
one has 
\bq\label{2.7}
\gd g_{\mu\nu}=2g_{\mu\nu}\gl;\,\,\gd\gb =-\gb\gl;\,\,\gd w_{\mu}=\pp_{\mu}\gl
\end{equation}
Putting these in \eqref{2.5} and integrating by parts gives then
\bq\label{2.8}
\gd I_M=16\pi\int[T-\na_{\gl}J^{\gl}-B\gb]\gl\sqrt{-g}d^4x
\end{equation}
Given the invariance of \eqref{2.5} under WGT one has from \eqref{2.8}
$({\bf 2K})\,\,T=\na_{\gl}J^{\gl}+\gb B$ and in view of ({\bf 2E}) this leads to
$({\bf 2L})\,\,T=\gb B$.  Further using ({\bf 2E}) and ({\bf 2L}) in ({\bf 2J})
yields $({\bf 2M})\,\,\na_{\gl}T^{\gl}_{\mu}-T(\pp_{\mu}\gb/\gb)=-J^{\gl}W_{\mu\gl}$
which is identical with the equations of motion for the matter ({\bf 2F}).  Going back to
the field equations \eqref{2.3} one contracts to get 
\bq\label{2.9}
G_{\gs}^{\gs}=-\frac{8\pi}{\gb^2}T+\frac{6}{\gb}\na^{\gl}\na_{\gl}\gb-4\gL\gb^2
\end{equation}
Putting ({\bf 2L}) into \eqref{2.9} and using $G_{\gs}^{\gs}=-R_{\gs}^{\gs}=R$ one obtains again the equation for the $\gb$ field \eqref{2.4}.  Consequently 
\eqref{2.4} is rather a corollary rather than an independent field equation and one
is free to choose the gauge function - this freedom indicates the gauge covariant
nature of the Weyl Dirac theory.
\\[3mm]\indent
We go now to the third paper in Israelit \cite{i2} and look at 
integrable Weyl-Dirac (Int-W-D) theory where $w_{\mu}=
\pp_{\mu}w$ and $W_{\mu\nu}=0$.  Thus the action in \eqref{2.1} has no $W^{\gl\mu}
W_{\gl\mu}$ term and one can introduce an ad hoc gradient $({\bf 2N})\,\,b_{\mu}=
\pp_{\mu}\gb/\gb$ with $W=w+b$ and $W_{\mu}=w_{\mu}+b_{\mu}$ (W will be gauge
invariant).  Replace now $k-6=16\pi\hat{\gs}$ and varying the action in \eqref{2.1}
with respect to w yields $({\bf 2O})\,\,2\na_{\nu}(\hat{\gs}\gb^2W^{\nu})=S$ where
S is called a Weyl scalar charge $({\bf 2P})\,\,16\pi S=\gd L_M/\gd w$.  Note here
that \eqref{2.3} can be rewritten as
\bq\label{2.11}
G^{\nu}_{\mu}=-8\pi\frac{T^{\mu}_{\nu}}{\gb^2}+16\pi\hat{\gs}\left[W^{\nu}W_{\nu}
-\frac{1}{2}\gd^{\nu}_{\mu}W^{\gs}W_{\gs}\right]+
\end{equation}
$$+2(\gd^{\nu}_{\mu}\na_{\gs}b^{\gs}-\na_{\mu}b^{\nu})+2b^{\nu}b_{\mu}+
\gd^{\nu}_{\mu}b^{\gs}b_{\gs}-\gd^{\nu}_{\mu}\gb^2\gL$$
and \eqref{2.4} as
\bq\label{2.12}
R^{\gs}_{\gs}+k(\na_{\gs}b^{\gs}+b^{\gs}b_{\gs})=16\pi\hat{\gs}(w^{\gs}w_{\gs}-
\na_{\gs}w^{\gs})+4\gb^2\gL+8\pi\gb^{-1}B
\end{equation}
B is defined as before and looking at CT and WGT transformations as above leads to
\bq\label{2.13}
\na_{\gl}T^{\gl}_{\mu}-Sw_{\mu}-\gb Bb_{\mu}=0;\,\,S+T-\gb B=0\Rightarrow
\na_{\gl}T^{\gl}_{\mu}-Tb_{\mu}=SW_{\mu}
\end{equation}
Going back to \eqref{2.11} one can introduce the energy momentum density tensor
of the $W_{\mu}$ field
\bq\label{2.14}
8\pi\gT^{\mu\nu}=16\pi\hat{\gs}\gb^2\left[\frac{1}{2}g^{\mu\nu}W^{\gl}W_{\gl}-W^{\mu}W^{\nu}\right]
\end{equation}
Making use of ({\bf 2O}) one obtains $({\bf 2Q})\,\,\na_{\gl}\gT^{\gl}_{\mu}-\gT b_{\mu}
=-SW_{\mu}$ and from \eqref{2.13} and ({\bf 2Q}) there results $({\bf 2R})\,\,
\na_{\gl}(T^{\gl}_{\mu}+\gT^{\gl}_{\mu})-(T+\gT)b_{\mu}=0$.  This gives a mini-survey
of the Int-W-D framework.  A fascinating W-D cosmology was developed in
\cite{i1,i2} but we omit this here (cf. also \cite{c16,c18,dt,d2,schz,sczh}).  There are
many general theories of relativity involving spin, torsion, nonmetricity, etc. and 
we refer to \cite{bl,h1,p1} for information (with apologies for omissions). 

\section{CONNECTIONS TO ELECTROMAGNETISM}
\renewcommand{\theequation}{3.\arabic{equation}}
\setcounter{equation}{0}

We go first to \cite{i1,i2,i4} and recall the Maxwell equations in the form
\bq\label{3.1}
\na\cdot {\bf E}=4\pi\rho;\,\,\na\times{\bf H}=4\pi{\bf j}+\pp_t{\bf E};\,\,
\na\times{\bf E}=-\pp_t{\bf H};\,\,\na\cdot{\bf H}=0
\end{equation}
The asymmetry regarding electric and magnetic currents is seen more clearly
in writing
\bq\label{3.2}
\na_{\nu}F^{\mu\nu}=4\pi J^{\mu};\,\,\na_{\nu}\tl{F}^{\mu\nu}=0;\,\,\tl{F}^{\mu\nu}
=-\frac{\gep^{\mu\nu\ga\gb}}{2\sqrt{-g}}F_{\ga\gb}
\end{equation}
where $\gep^{\mu\nu\ga\gb}$ is the completely antisymmetric Levi-Civita 
symbol.  From this one can write $({\bf 3A})\,\,F_{\mu\nu}=\na_{\nu}A_{\mu}-
\na_{\mu}A_{\nu}$ which in current free regions can be written in Lorentz
gauge as $({\bf 3B})\,\,\na_{\nu}\na^{\nu}A^{\mu}+A^{\nu}R_{\nu}^{\mu}=0$.
In order to build up a theory admitting intrinsic magnetic and electric currents
and massive photons one asumes $\gs=k-6\ne 0$ which for $k<6$ leads to a
Proca field that can be interpreted as an ensemble of massive bosons 
with spin one (i.e. massive photons).  Thus take a symmetric metric $g_{\mu\nu}=
g_{\nu\mu}$, a Weyl vector $w_{\mu}$, a Dirac gauge function $\gb$, and a 
torsion tensor $\bar{\gG}^{\gl}_{[\mu\nu]}$ as in \eqref{1.18} with contorsion tensor
\bq\label{3.3}
C^{\gl}_{\mu\nu}=\bar{\gG}^{\gl}_{[\mu\nu]}-
g^{\gl\gb}g_{\gs\mu}\bar{\gG}^{\gs}_{[\gb\nu]}-g^{\gl\gb}g_{\gs\nu}
\bar{\gG}^{\gs}_{[\gb\mu]}
\end{equation}
and under WGT one has
\bq\label{3.4}
g_{\mu\nu}\to\tl{g}_{\mu\nu}=e^{2\gl}g_{\mu\nu};\,\,w_{\mu}\to
\tl{w}_{\mu}=w_{\mu}+\pp_{\mu}\gl;\,\,\gb\to\tl{\gb}=e^{-\gl}\gb
\end{equation}
and one assumes $({\bf 3C})\,\,\bar{\gG}^{\gl}_{[\mu\nu]}\to\tl{\gG}^{\gl}_
{[\mu\nu]}=\bar{\gG}^{\gl}_{\mu\nu}$ with a curvature formula as in \eqref{1.15}, namely
\bq\label{3.5}
K^{\gl}_{\mu\nu\gs}=-\pp_{\gs}\bar{\gG}^{\gl}_{\mu\nu}+\pp_{\nu}\bar{\gG}^{\gl}_
{\mu\gs}-\bar{\gG}^{\ga}_{\mu\nu}\bar{\gG}^{\gl}_{\ga\gs}+\bar{\gG}^{\ga}_
{\mu\gs}\bar{\gG}^{\gl}_{\ga\nu}
\end{equation}
Further there is a formula \eqref{1.19} describing geometrical properties 
invoked via torsion where (via $\bar{\na}_{\gl}g_{\mu\nu}=-2g_{\mu\nu}w_{\gl}$)
one can express the torsional curvature term as $({\bf 3D})\,\,-2\bar{\na}_{\ga}
T_{\mu}\bar{\gG}^{\ga}_{[\nu\gs]}\sim a\bar{\na}_{\ga}W_{\mu\gl}\bar{\gG}^{\ga}_
{[\nu\gs]}$ where $W_{\mu\nu}$ is the Weyl length curvature tensor.  The
action integral \eqref{2.1} is then replaced by a three line expression
(cf. \cite{i1}, p. 63) with
dynamical variables  
$\bar{\gG}^{\gl}_{[\mu\nu]},\,\,w_{\mu},\,\,g_{\mu\nu},$ and $\gb$ (a dependence of $L_M$ on additional matter field variables is not excluded).
\\[3mm]\indent
Now varying $w_{\mu}$ in this action gives
\bq\label{3.6}
\na_{\nu}\left[W^{\mu\nu}-2\na_{\ga}\bar{\gG}^{\ga[\mu\nu]}\right]=\frac{\gb^2}{2}
(k-6)W^{\mu}+2\gb^2\bar{\gG}^{\ga[\mu}_{\,\,\,\ga]}+4\pi J^{\mu}
\end{equation}
where $({\bf 3E})\,\,W_{\mu}=w_{\mu}+\pp_{\mu}log(\gb)$ and under WGT 
$({\bf 3F})\,\,\tl{W}_{\mu}=W_{\mu}$ with $16\pi J^{\mu}=\gd L_M/\gd w_{\mu}$
($W_{\mu}$ is called the gauge invariant Weyl connection vector
and $J^{\mu}=0$ if $L_M$ does not depend on $w_{\mu}$).
Variation of the action with respect to $\bar{\gG}^{\gl}_{[\mu\nu]}$ gives another field
equation (cf. Israelit \cite{i1}, p. 64)
involving
$16\pi\gO^{[\mu\nu]}_{\gl}=\gd L_M/\gd\bar{\gG}^{\gl}_{[\mu\nu]}$ whose contraction 
yields
\bq\label{3.7}
\na_{\nu}W^{\mu\nu}=-3\gb^2W^{\mu}+2\gb^2\bar{\gG}^{\nu[\mu}_{\,\,\,\nu]}-4\pi\gO_{\nu}^
{[\mu\nu]}
\end{equation}
One regards $J^{\mu}$ as the electric current vector and the magnetic current
density vector will be expressed in terms of $\gO_{\gl}^{[\mu\nu]}$.  One has two
conservation laws following from \eqref{3.6} and \eqref{3.7}, namely
\bq\label{3.8}
(k-6)\na_{\mu}(\gb^2W^{\mu})+8\pi \na_{\mu}J^{\mu}+4\na_{\mu}(\gb^2\bar{\gG}^
{\nu[\mu}_{\,\,\,\nu]})=0
\end{equation}
\bq\label{3.9}
3\na_{\mu}(\gb^2W^{\mu})+4\pi\na_{\mu}\gO^{[\mu\nu]}_{\nu}-2\na_{\mu}(\gb^2
\bar{\gG}^{\nu[\mu}_{\,\,\,\nu]})=0
\end{equation}
Further calculation leads to the introduction of a strength tensor for the EM fields
\bq\label{3.10}
\Phi_{\mu\nu}=W_{\mu\nu}-2\na_{\ga}\bar{\gG}^{\ga}_{[\mu\nu]}\equiv
\na_{\nu}W_{\mu}-\na_{\mu}W_{\nu}-2\na_{\ga}\bar{\gG}^{\ga}_{[\mu\nu]}
\end{equation}
allowing one to write \eqref{3.6} in the form
\bq\label{3.11}
\na_{\nu}\Phi^{\mu\nu}=\frac{1}{2}\gb^2(k-6)W^{\mu}+2\gb^2\bar{\gG}^{\ga[\mu}_{\,\,\,\ga]}
+4\pi J^{\mu}
\end{equation}
(note $\Phi_{\mu\nu}=0$ if $W_{\mu\nu}=0$ and $\bar{\gG}^{\ga}_{[\mu\nu]}=0$).
The dual field tensor is now $({\bf 3G})\,\,\tl{\Phi}^{\mu\nu}=-(1/2\sqrt{-g})\gep^{\mu\nu\ga\gb}\Phi_{\ga\gb}$ and one can write
\bq\label{3.12}
\na_{\gl}\Phi_{\mu\nu}+\na_{\nu}\Phi_{\gl\mu}+\na_{\mu}\Phi_{\nu\gl}=4\pi(\gO_{\mu[\nu\gl]}+\gO_{\gl[\mu\nu]}+\gO_{\nu[\gl\mu]})\equiv 4\pi\gT_{\mu\nu\gl}
\end{equation}
with $\gO_{\gl[\mu\nu]}\equiv g_{\mu\ga}g_{\nu\gb}\gO_{\gl}^{[\ga\gb]}$.  This all leads
then to $({\bf 3H})\,\,\na_{\nu}\tl{\Phi}^{\mu\nu}=4\pi L^{\mu}$ where $L^{\mu}=
-(1/6\sqrt{-g})\gep^{\mu\gl\nu\gs}\gT_{\gl\nu\gs}$ with $\na_{\mu}L^{\mu}=0$.  Thus one
obtains two equations \eqref{3.11} and ({\bf 3H}) for $\Phi_{\mu\nu}$ and
$\tl{\Phi}_{\mu\nu}$.  The fields are created by intrinsic magnetic and electric currents
and for $k\ne 6$ a Proca term appears in the $\Phi_{\mu\nu}$ equation.
\\[3mm]\indent
This is developed to considerable extent in Israelit \cite{i1,i2,i4} and the torsion plays a 
crucial role in that one can generate a dual field tensor having a nonvanishing 
divergence (making possible magnetic currents).  Indeed one can define
$({\bf 3I})\,\,\bar{\gG}^{\gl[\mu\nu]}=-(1/\sqrt{-g})\gep^{\gl\mu\nu\gs}V_{\gs}$ leading to
\bq\label{3.13}
\na_{\nu}\Phi^{\mu\nu}=\na_{\nu}\na^{\nu}W^{\mu}-\na_{\nu}\na^{\mu}W^{\nu}=
\frac{k-6}{2}\gb^2W^{\mu}+4\pi J^{\mu};
\end{equation}
$$\na_{\nu}\tl{\Phi}^{\mu\nu}=\na_{\nu}\na^{\nu}V^{\mu}-\na_{\nu}\na^{\mu}V^{\nu}
=-2\pi L^{\mu}$$
Working in the Einstein gauge $\gb=1$ one has a Proca equation $({\bf 3J})\,\,
\na_{\nu}\na^{\nu}w^{\mu}+\gk^2w^{\mu}=0$ ($\gk^2=(1/2)(6-k)$) and in the
absence of magnetic fields $V^{\mu}$ a massless photon 
is implied via $\gk=0$ (cf. \cite{as,h1,h2,h3} for other points of view).

\section{DIVERSE DEVELOPMENTS IN WEYL GEOMETRY}
\renewcommand{\theequation}{4.\arabic{equation}}
\setcounter{equation}{0}

We gather in this section some material from the literature in an attempt to clarify
and unify various approaches (cf. \cite{au,av,aw,bl,
c18,dt,d2,fjm,fu,h1,l1,m1,pe,q1,r1,sa,
schz,sczh,s35,ti,w1}).  We coordinate notation by noting that the notation for the
Weyl connection differs in many of these references but they are all equivalent
modulo a factor of $\pm 1$ or $\pm 1/2$ in the definition of $w_{\mu}$.  Thus
e.g. \cite{i1} $\equiv$ \cite{s35}$\equiv$ \cite{l1}, \cite{sczh} $\equiv$ \cite{w1} $\equiv$ -\cite{i1},
\cite{dt} $\equiv$ \cite{fu} $\equiv$ -\cite{au} $\equiv$ -\cite{bl} $\equiv$ (1/2)\cite{i1}.
There are various approaches to Weyl geometry and we only try to build on the
framework already developed in Section 1-2.  Thus go to \cite{l1} which has a compatible
point of view and start with transformations $({\bf 4A})\,\,g_{\mu\nu}\to exp(2\gl)
g_{\mu\nu}$ with length changes $({\bf 4B})\,\,\ell\to exp(\gl)\ell$; then the relation 
\eqref{1.10} holds under Weyl transformations $w_{\mu}\to w_{\mu}+\pp_{\mu}\gl$.  If
$\pp_{\mu}w_{\nu}-\pp_{\nu}w_{\mu}=0$ there is a Weyl transformation reducing
$w_{\mu}$ to zero and the space is Riemannian (or semi-Riemannian).  
Here $\hat{\gG}^{\gl}_{\mu\nu}$ is given by \eqref{1.7} and on has curvature $K_{\nu\mu}$
from \eqref{1.17} with K as in ({\bf 1I}).  There is a nice discussion in \cite{l1} of
attempts to develop EM in the Weyl theory and its difficulties.  In particular one problem
is that the Weyl vector does not seem to couple to spinors (see the last chapter in \cite{l1}).  The Einstein-Schr\"odinger approach to EM is also discussed along with Brans-Dicke
and Jordan theories.  In terms of conformal geometry one considers conformally 
equivalence via $\tl{g}_{\mu\nu}=exp(2\gl)g_{\mu\nu}$ (or what is the same $\tl{ds}=
exp(\gl)ds$) with corresponding connection $\hat{\gG}^{\gl}_{\mu\nu}$ as in \eqref{1.7}
with $w_{\mu}\to -\pp_{\mu}\gl$ (integrable Weyl geometry).  One gets then (in 4-D)
\bq\label{4.1}
\tl{R}_{\nu\rho}=R_{\nu\rho}+g_{\nu\rho}\bx\gl+2(\na_{\rho}\na_{\nu}\gl+g_{\nu\rho}
\pp_{\gs}\gl\pp^{\gs}\gl -\pp_{\nu}\gl\pp_{\rho}\gl)
\end{equation}
leading to
\bq\label{4.2}
\tl{R}=e^{-2\gl}[R+6\bx \gl+6\pp_{\gs}\gl\pp^{\gs}\gl]
\end{equation}
where $({\bf 4C})\,\,\bx\gl=(1/\sqrt{-g})\pp_{\mu}(\sqrt{-g}g^{\mu\nu}\pp_{\nu}\gl)$ as usual.
Equation \eqref{4.2} has an alternate form since in 4-D the transformation law for a 
curvature scalar is $({\bf 4D})\,\,\tl{R}=exp(-2\gl)[R+2(\bx \gl+\pp_{\mu}\gl\pp^{\mu}\gl)]$
leading to 
\bq\label{4.3}
\tl{R}=e^{-2\gl}\left[R+6e^{-\gl}\bx e^{\gl}\right]
\end{equation}
in place of \eqref{4.2}.  One shows also that for $\tl{\phi}=exp(-\gl)\phi$ there results
\bq\label{4.4}
\stackrel{\tl{}}{\bx}\tl{\phi}+\frac{1}{6}\tl{R}\tl{\phi}=e^{-3\gl}\left[\bx\phi+\frac{1}{6}R\phi\right]
\end{equation}
so $\bx\phi+(1/6)R\phi=0$ is conformally invariant.
\\[3mm]\indent
There is also information on scalar-tensor theories in \cite{l1} (cf. also \cite{fjm} for
a string oriented approach).  We recall first \eqref{1.10} $\hat{\na}g_{\mu\nu}=2g_{\mu\nu}
w_{\gl}$ or in the notation of \cite{l1} $({\bf 4E})\,\,\hat{\na}_{\rho}g_{\mu\nu}=2g_{\mu\nu}w_{\rho}$.  This is invariant under WGT $({\bf 4F})\,\,g_{\mu\nu}\to exp(2\gl)g_{\mu\nu}$
and $w_{\mu}\to w_{\mu}+\pp_{\mu}\gl$.  A physical quantity $X$ with dimension $L^N$
can be assigned a transformation law $X\to exp(N\gl)X$ under a conformal map.  Thus
conformal maps are spacetime dependent redefinitions of the length unit.  The metric cannot be used for raising and lowering indices since it does not commute with the Weyl
covariant derivative.  However in \cite{l1} (cf. also \cite{d1}) one defines a new covariant derivative which is 
covariant for unit transformations as well as coordinate transformations.  For this let
$\dot{\na}$ denote any covariant differentiation for some unspecified $\dot{\gG}^{\mu}_
{\nu\rho}$ (e.g. $\hat{\na}$ and $\hat{\gG}{\mu}_{\nu\rho}$).  Then if $X$ is any tensor density with dimension number N the transformation
law for the covariant derivative under a unit transformation as above will be
$({\bf 4G})\,\,\dot{\na}_{\rho}X\to exp(N\gl)(\dot{\na}_{\rho}X+N\pp_{\rho}\gl X)$.
In order to eliminate the final term in ({\bf 4G}) one introduces a conformally covariant
derivative $({\bf 4H})\,\,\ddot{\na}_{\rho}X=\dot{\na}_{\rho}X-N\phi_{\rho}X$ where
$\phi_{\rho}$ is the Weyl vector with transformation law ({\bf 4F}); this has the 
transformation law $({\bf 4I})\,\,\ddot{\na}_{\rho}X\to exp(N\gl)\ddot{\na}_{\rho}X$.
To determine the
components of the affine connection one imposes the restriction $({\bf 4J})\,\,
\ddot{\na}_{\rho}g_{\mu\nu}=0$ and since $N=2$ for $g_{\mu\nu}$ this leads to
\bq\label{4.5}
\ddot{\gG}^{\rho}_{\mu\nu}=\gG^{\rho}_{\mu\nu}+g_{\mu\nu}\phi^{\rho}-\gd_{\mu}^{\rho}
\phi_{\nu}-\gd_{\nu}^{\rho}\phi_{\mu}
\end{equation}
and this is precisely the Weyl affine connection $\hat{\gG}^{\rho}_{\mu\nu}$ of 
\eqref{1.7} ($\ddot{\na}_{\mu}$ is a ``conformalized" covariant Weyl derivative). 
Because of ({\bf 4J}) conformally covariant differentiation commutes with
the raising and lowering of vector indices and this new formalism is fully equivalent to
that of Weyl since ({\bf 4J}) is precisely the same equation as ({\bf 4E}).  Recall that
the Christoffel symbols are defined via $\na_{\rho}g_{\mu\nu}=0$.  Now $\phi_{\mu}$ 
cannot be assigned a conformally covariant derivative because of its peculiar transformation law under unit transformations; it does however have a conformally
covariant curl
\bq\label{4.6}
\phi_{\mu\nu}=\pp_{\mu}\phi_{\nu}-\pp_{\nu}\phi{\mu}=\na_{\mu}\phi_{\nu}-\na_{\nu}
\phi_{\mu}
\end{equation}
which is a tensor with zero dimension number.  One can also write
for a vector of dimension number (or Weyl weight) N and scalar density $w$
\bq\label{4.7}
\ddot{\na}_{\rho}\ddot{\na}_{\nu}X_{\mu}-\ddot{\na}_{\nu}\ddot{\na}_{\rho}X_{\mu}=
P^{\gl}_{\,\mu\nu\rho}X_{\gl}+(N-4w)\phi_{\nu\rho}X_{\mu}
\end{equation}
and for a contravariant vector 
\bq\label{4.8}
\ddot{\na}_{\rho}\ddot{\na}_{\nu}X^{\mu}-\ddot{\na}_{\nu}\ddot{\na}_{\rho}X^{\mu}=
-P^{\mu}_{\,\gl\nu\rho}+(N-4w)\phi_{\nu\rho}X^{\mu}
\end{equation}
(note if $A\in GL(N)$ then $\tl{y}=|A|^wy$ is a representation and $y$ is said to be
a scalar density of weight $w$).
The tensor $P^{\gl}_{\,\mu\nu\rho}$ is the curvature tensor constructed from 
\eqref{4.5} which means it is the Weyl curvature $K^{\gl}_{\mu\nu\rho}$; it is 
conformally invariant.  The generalization to more general tensors follows the pattern
(for a skew symmetric tensor density)
\bq\label{4.9}
2\ddot{\na}_{\rho}\ddot{\na}_{\nu}X^{\nu\rho}=-P^{\nu}_{\,\,\,\gl\nu\rho}X^{\gl\rho}-
P^{\rho}_{\,\,\gl\nu\rho}X^{\nu\gl}+(N-4w)\phi_{\nu\rho}X^{\nu\rho}=
\end{equation}
$$=(P_{\nu\rho}-P_{\rho\nu})X^{\nu\rho}+(N-4w)\phi_{\nu\rho}X^{\nu\rho}=
(4+N-4w)\phi_{\nu\rho}X^{\nu\rho}$$
(one has used \eqref{1.17} here).  For $\phi^{\mu\nu}$ itself (where $w=0$ and 
$N=-4$) one has $\ddot{\na}_{\rho}\ddot{\na}_{\nu}\phi^{\nu\rho}=0$.
\\[3mm]\indent
The obvious choice of Lagrangian for the $\phi_{\mu}$ field is, by analogy with EM,
$({\bf 4K})\,\,-(1/4)\sqrt{-g}(1/f^2)\phi_{\mu\nu}\phi^{\mu\nu}$ where $f$ is a dimensionless
coupling constant (the physical vector field is then actually $\chi_{\mu}$ where
$\phi_{\mu}=f\chi_{\mu}$ and $f$ represents the strength of the coupling of $\chi_{\mu}$
to other fields).  One wants now to build a Lagrangian all of whose parts have zero 
dimension number (such as ({\bf 4K})).  The mass is an inverse length with $N=-1$ so
one writes $m=\gs \mu$ whre $\mu$ is a dimensionless constant and the Lagrangian
density for the associated field $\gs$ is then taken to be $({\bf 4L})\,\,-(1/2)
\sqrt{-g}\ddot{\na}_{\mu}\gs\ddot{\na}^{\mu}\gs$ where $\ddot{\na}_{\mu}\gs=
\pp_{\mu}\gs+\phi_{\mu}\gs$.  One then shows that a Dirac spinor field is not coupled
to the field $\phi_{\mu}$ so that $\phi_{\mu}$ cannot be interpreted as EM
(see \cite{l1} for details).
\\[3mm]\indent
{\bf REMARK 4.1.}
Going to Canuto et al \cite{c19} one looks at Einstein units where $({\bf 4M})\,\,
\bar{G}_{\mu\nu}=\bar{R}_{\mu\nu}-(1/2)\bar{g}_{\mu\nu}\bar{R}=-8\pi\bar{T}_{\mu\nu}+
\bar{\gL}\bar{g}_{\mu\nu}$ and sets $\bar{g}_{\mu\nu}=\ga^2g_{\mu\nu}$ (conformal
frame) to get
\bq\label{4.10}
\bar{G}_{\mu\nu}=G_{\mu\nu}+\frac{2\na_{\nu}\gb_{\mu}}{\gb}-\frac{4\gb_{\mu}\gb_{\nu}}
{\gb^2}-g_{\mu\nu}\left(2\frac{\na_{\gl}\gb^{\gl}}{\gb}-\frac{\gb^{\gl}\gb_{\gl}}{\gb^2}\right)
\end{equation}
($\na$ here refers to $g_{\mu\nu}$ and $\bar{\gL}\bar{g_{\mu\nu}}=\gL g_{\mu\nu}$ for
$\gL=\ga^2\bar{\gL}$).  The right side of \eqref{4.10} is form invariant (i.e. $g_{\mu\nu}=
\chi^2g'_{\mu\nu}$ leads to \eqref{4.10} with $\ga\to\ga\chi$ and covariant derivative
$\na'$ referring to $g'_{\mu\nu}$.  The spacetime underlying the ensuing Einstein equations
\bq\label{4.11}
G_{\mu\nu}+2\frac{\na_{\nu}\gb_{\mu}}{\gb}-4\frac{\gb_{\mu}\gb_{\nu}}{\gb^2}-g_{\mu\nu}
\left(2\frac{\na_{\gl}\gb^{\gl}}{\gb}-\frac{\gb^{\gl}\gb_{\gl}}{\gb^2}\right)=
-8\pi T_{\mu\nu}+\gL g_{\mu\nu}
\end{equation}
is an integrable Weyl  (IW) manifold (i.e. the Weyl vector $\phi_{\mu}$ satisfies
$\na_{\nu}\phi_{\mu}-\na_{\mu}\phi_{\nu}=0$ with $\phi_{\mu}=\pp_{\mu}\Phi$).
This means that the spacetime is conformally equivalent to a Riemannian space but
$\phi_{\mu}$ does have physical significance.  For the Weyl geometry we have as usual
the connection $\hat{\gG}^{\gl}_{\mu\nu}$ as in \eqref{1.7}
with curvature K as in ({\bf 1V}).  Thus the basic theory corresponds to our previous 
development.$\hfill\bs$

\section{BOHMIAN QUANTUM GRAVITY}
\renewcommand{\theequation}{5.\arabic{equation}}
\setcounter{equation}{0}

We go here to work of A. and F. Shojai and M. Golshani \cite{s35,s17,ss3} which has had a profound effect on our thinking
(cf. \cite{c1,c3,c4,c5} and see also \cite{an,au,av,aw,b9,c16,c17,d1,fa,fe,pa,sa,ti,w1,w2}).
We follow the notation of A. and F. Shojai \cite{s35} which is visibly compatible with Sections 1-3 but 
some adjustment seems necessary in the notation (we will therefore also
indicate the equations of \cite{d1} and \cite{i3,r1} below 
in an attempt to unify all approaches -
for brevity we omit the basic points of \cite{c17}).
Going then to \cite{s35} one refers to \cite{ad,d1}
and begins with $({\bf 5A})\,\,\gd\ell=\phi_{\mu}\gd x^{\mu}\ell$ or equivalently
$\ell=\ell_0exp(\int\phi_{\mu}dx^{\mu})$.  The length change corresponds to a metric
change $({\bf 5B})\,\,g_{\mu\nu}\to exp(2\int\phi_{\mu}dx^{\mu})g_{\mu\nu}$ and one
arrives at a Weyl affine connection as before, namely $\hat{\gG}^{\mu}_{\nu\gl}\sim
\ddot{\gG}^{\mu}_{\nu\gl}$.  Given a transformation $({\bf 5C})\,\,\phi_{\mu}\to
\phi'_{\mu}=\phi_{\mu}+\pp_{\mu}\gL$ one has $\gd\ell\to \gd\ell+(\pp_{\mu}\gL)\gd x^{\mu}
\ell$.  The curl of the Weyl vector is $F_{\mu\nu}=\pp_{\mu}\phi_{\nu}-\pp_{\nu}\phi_{\mu}$
has features similar to and EM field strength but as noted in \cite{l1} should not be
identified with EM.  Now the Weyl-Dirac action is taken as
\bq\label{5.1}
A=\int d^4x\sqrt{-g}(F_{\mu\nu}F^{\mu\nu}-\gb^2K+(\gs+6)\ddot{\na}_{\mu}\gb\ddot{\na}^{\mu}
\gb+L_M)
\end{equation}
where $\gb$ is a scalar field of weight $-1$.  The covariant derivative here is written as
$({\bf 5D})\,\,\ddot{\na}_{\mu}X=\hat{\na}_{\mu}X-N\phi_{\mu}X$ (discussed below).
\\[3mm]\indent
{\bf REMARK 5.1.}
Let us group together some versions of the action.  First from \cite{l1} in ({\bf 4H})
one has $\ddot{\na}_{\rho}X=\dot{\na}_{\rho}X-N\phi_{\rho}X$ which agrees with e.g. \cite{i1} and ({\bf 5D})
should be tuned to this.  Then we want to compare A in \eqref{5.1}
with I of \eqref{2.1} (recalling $(\bgs)$ and $(\bgs\bgs)$ after \eqref{2.1} and the 
reduction $\gb^2\na_{\gl}\phi^{\gl}\to -2\gb\phi^{\gl}\pp_{\gl}\gb$ after integration.  We 
write for comparison some action formulas to compare with \eqref{5.1}.  Thus
\eqref{2.1} gives (cf. \cite{i1})
\bq\label{5.2}
I=\int [W^{\gl\mu}W_{\gl\mu}-\gb^2R+\gs\gb^2w^{\gl}w_{\gl}+(\gs+6)\pp_{\gl}\gb
\pp^{\gl}\gb+
\end{equation}
$$+2\gs\gb\pp_{\gl}\gb w^{\gl}+2\gL\gb^4+L_M]\sqrt{-g}d^4x$$
while \eqref{3.14} (from \cite{i3}) is 
\bq\label{5.3}
L_D=W_{\mu\nu}W^{\mu\nu}-\gb^2R+k\pp_{\mu}\gb\pp^{\mu}\gb+(k-6)(2\gb w^{\mu}
\gb_{\mu}+\gb^2w_{\mu}w^{\mu})+2\gL\gb^4
\end{equation}
On the other hand from \cite{r1} one has $({\bf 5E})\,\,{}^*\na_{\nu}A_{\mu}=
\hat{\na}_{\nu}A_{\mu}-N\phi_{\nu}A_{\mu}\equiv \ddot{\na}_{\nu}A_{\mu}$ as a 
comformalized Weyl derivative (cf. ({\bf 4H}) with $({\bf 5E})\,\,\hat{\na}_{\gs}
g_{\mu\nu}=2g_{\mu\nu}\phi_{\gs}\equiv\ddot{\na}_{\gs}g_{\mu\nu}=0$). 
Rosen \cite{r1} also discusses connections $({\bf 5F})\,\,\tl{\gG}^{\gs}_{\mu\nu}=\hat{\gG}^{\gs}_{\mu\nu}
+\gd_{\mu}^{\gs}\phi_{\nu}$ with $\tl{\gG}_{\gs}g_{\mu\nu}=0$ and acting on ``standard"
vectors ($\Pi(\xi_{\mu})=1$ or $\Pi(\xi^{\mu})=-1$) one has e.g. $({\bf 5G})\,\,
\tl{\na}_{\nu}\xi_{\mu}=\ddot{\na}_{\nu}\xi_{\mu}$  One notes that $\tl{\gG}^{\gs}_{\mu\nu}$ involves torsion.  Rosen then takes for an in-invariant action ($F\sim W$)
\bq\label{5.4}
{\mf A}=F_{\mu\nu}F^{\mu\nu}-\gb^2K+k\ddot{\na}_{\mu}\gb\ddot{\na}^{\mu}\gb
+2\gL\gb^4+L_M
\end{equation}
where $\ddot{\na}_{\mu}\gb\ddot{\na}^{\mu}\gb=(\pp_{\mu}\gb+\phi_{\mu}\gb)
(\pp^{\mu}\gb+\phi^{\mu}\gb)$ (recall $N(\gb)=-1$) so that $({\bf 5H})\,\,
k\ddot{\na}_{\mu}\gb\ddot{\na}^{\mu}\gb=\pp_{\mu}\gb\pp^{\mu}\gb +2\phi_{\mu}\gb
\pp^{\mu}\gb+\gb^2\phi_{\mu}\phi^{\mu}$.  Hence, setting $k-6=\gs$ one has
(cf. \eqref{2.1})
\bq\label{5.5}
{\mf A}=F_{\mu}F^{\mu\nu}+k[\pp_{\mu}\gb\pp^{\mu}\gb+2\phi_{\mu}\gb\pp^{\mu}\gb+
\gb^2\phi_{\mu}\phi^{\mu}]+
\end{equation}
$$+2\gL\gb^4+L_M-\gb^2R-6\gb^2\phi^{\gl}\phi_{\gl}
-6\cdot 2\gb\phi^{\gl}\pp_{\gl}\gb
=F_{\mu\nu}F^{\mu\nu}+$$
$$+(\gs+6)\pp_{\mu}\gb\pp^{\mu}\gb+2\gL\gb^4+L_M-\gb^2R+\gs\phi_{\mu}\phi^{\mu}
+2\gs\gb\phi^{\mu}\pp_{\mu}\gb$$
as in \cite{r1} (and this agrees, not surprisingly, with \cite{i1} and \eqref{2.1}).
Thus we see that \eqref{5.1} is the same as \eqref{5.4} where we only need to revise
the notation of \cite{s35} to be $({\bf 5I})\,\,X_{;\mu}=\ddot{\na}_{\mu}X={}^W\na_{\mu}X
-N\phi_{\mu}X=\hat{\na}_{\mu}X-N\phi_{\mu}X$ identifying ${}^W\na_{\mu}X$ with
$\hat{\na}_{\mu}X$ and $X_{;\mu}=\ddot{\na}_{\mu}X$.$\hfill\bs$
\\[3mm]\indent
We continue now with our sketch from F. and A. Shojai \cite{s35} 
and will extract more or
less freely in order to accurately transmit the thrust of the arguments.
Thus one has
equations of motion
\bq\label{5.6}
G^{\mu\nu}=-\frac{8\pi}{\gb^2}\left(T^{\mu\nu}+M^{\mu\nu}\right)+\frac{2}{\gb}
\left(g^{\mu\nu}
\hat{\na}^{\ga}\hat{\na}_{\ga}\gb-\hat{\na}^{\mu}\hat{\na}^{\nu}\gb\right)+
\end{equation}
$$+\frac{1}{\gb^2}\left(4\na^{\mu}\gb\na^{\nu}\gb-g^{\mu\nu}\na^{\ga}\gb\na_{\ga}\gb\right)+\frac{\gs}{\gb^2}\left(\ddot{\na}^{\mu}\gb\ddot{\na}^{\nu}\gb
-\frac{1}{2}g^{\mu\nu}\ddot{\na}^{\ga}\gb\ddot{\na}_{\ga}\gb\right)$$
\bq\label{5.7}
\hat{\na}_{\nu}F^{\mu\nu}=\frac{1}{2}\gs\left(\gb^2\phi^{\mu}+\gb\na^{\mu}\gb\right)
+4\pi J^{\mu};
\end{equation}
$$R=-(\gs+6)\frac{\stackrel{\hat{}}{\bx}\gb}{\gb}+\gs\phi_{\ga}\phi^{\ga}-\gs\hat{\na}^{\ga}
\phi_{\ga}+\frac{\psi}{2\gb}$$
where $({\bf 5J})\,\,M^{\mu\nu}=(1/4\pi)[(1/4)g^{\mu\nu}F^{\ga\gb}F_{\ga\gb}
-F^{\mu}_{\ga}F^{\nu\ga}$] and
\bq\label{5.8}
8\pi T^{\mu\nu}=\frac{1}{\sqrt{-g}}\frac{\gd\sqrt{-g}L_M}{\gd g_{\mu\nu}};\,\,
16\pi J^{\mu}=\frac{\gd L_M}{\gd \phi_{\mu}};\,\,\psi=\frac{\gd L_M}{\gd\gb}
\end{equation}
The equation of motion of matter and the trace of the energy momentum tensor
can be obtained from the invariance of action under coordinate and gauge
transformations yielding
\bq\label{5.9}
\hat{\na}_{\nu}T^{\mu\nu}-T\frac{\na^{\mu}\gb}{\gb}=J_{\ga}\phi^{\ga\mu}
-\left(\phi^{\mu}+\frac{\na^{\mu}\gb}{\gb}\right)\hat{\na}_{\ga}J^{\ga};\,\,
16\pi T-16\hat{\na}_{\mu}J^{\mu}-\gb\psi=0
\end{equation}
The first equation in \eqref{5.9} is a geometrical identity (Bianchi) and the 
second results from the non-independence of the field equations.  The Weyl
vector is being used as part of the geometry of spacetime and is in no way
confused with an EM field.  One will see that the Dirac field $\gb$ represents 
a quantum mass field and both the gravitational fields $g_{\mu\nu}$ and $\phi_{\mu}$ together with the quantum mass field $\gb$ determine the geometry 
of spacetime.  In some sense the gravitational fields determine the causal
structure of spacetime while the quantum effects give the conformal structure.
However this does not mean that the quantum effects have nothing to do with
the causal structure since it can act there via back-reaction terms in the field
equations.
The conformal factor is in fact a function of the dimensionless quantum potential
$({\bf 5K})\,\,Q=\ga(\bx\sqrt{\rho}/\sqrt{\rho}$ where $\ga=\hbar^2/m^2c^2$
and $\rho$ is an ensemble density of the system.  In fact one has a
quantum Hamilton-Jacobi (HJ) equation $({\bf 5L})\,\,\na_{\mu}S\na^{\mu}S
={\mf M}^2c^2=m^2c^2exp(Q)\sim m^2c^2(1+Q)$ (see here \cite{c3,n1} for
the exponential $exp(Q)$).  One can either consider the quantum mass field ${\mf M}$
and the classical metric or a classical mass and a quantum metric to obtain
the same results.  Thus different conformal frames are identical pictures of the
gravitational and quantum phenomena; one would feel different quantum forces
in different conformal frames.  This can be formulated into a conformal 
equivalence principle (cf. \cite{s35}) and the Weyl geometry provides a unified
geometrical framework for dealing with the gravitational and quantum forces.
One will have geodesic equations
\bq\label{5.10}
\frac{dx^{\mu}}{d\tau^2}+\gG^{\mu}_{\nu\gl}\frac{dx^{\nu}}{d\tau}\frac{dx^{\gl}}
{d\tau}=\frac{1}{{\mf M}}\left(g^{\mu\nu}-\frac{dx^{\mu}}{d\tau}\frac{dx^{\nu}}{d\tau}
\right)\na_{\nu}{\mf M}
\end{equation}
and we refer to \cite{s35} for more philosophy.
\\[3mm]\indent
One now simplifies the model by assuming that the matter Lagrangian does not depend on the Weyl vector so $J_{\mu}=0$; then the equations of motion are
\bq\label{5.11}
G^{\mu\nu}=-\frac{8\pi}{\gb^2}\left(T^{\mu\nu}+M^{\mu\nu}\right)+\frac{2}{\gb}
\left(g^{\mu\nu}\hat{\na}^{\ga}\hat{\na}_{\ga}\gb-\hat{\na}^{\mu}\hat{\na}^{\nu}\gb
\right)+
\end{equation}
$$+\frac{1}{\gb^2}\left(5\na^{\mu}\gb\na^{\nu}\gb-g^{\mu\nu}\na^{\ga}\gb
\na_{\ga}\gb\right)+\frac{\gs}{\gb^2}\left(\ddot{\na}^{\mu}\gb\ddot{\na}^{\nu}\gb
-\frac{1}{2}g^{\mu\nu}\ddot{\na}^{\ga}\gb\ddot{\na}_{\ga}\gb\right)$$
\bq\label{5.12}
\hat{\na}_{\nu}F^{\mu\nu}=\frac{1}{2}\gs(\gb^2\phi^{\mu}+\gb\na^{\mu}\gb);\,\,
R=-(\gs+6)\frac{\stackrel{\hat{}}{\bx}\gb}{\gb}+\gs\phi_{\ga}\phi^{\ga}-\gs\hat{\na}^{\ga}
\phi_{\ga}+\frac{\psi}{2\gb}
\end{equation}
\bq\label{5.13}
\hat{\na}_{\nu}T^{\mu\nu}-T\frac{\na^{\mu}\gb}{\gb}=0;\,\,16\pi T-\gb\psi=0
\end{equation}
Note here from \eqref{5.12} that $({\bf 5M})\,\,\hat{\na}_{\mu}(\gb^2\phi^{\mu}
+\gb\na^{\mu}\gb)=0$ so that $\phi_{\mu}$ is not independent of $\gb$. 
\\[3mm]\indent
Now for the quantum mass field one notes from \eqref{5.12} and \eqref{5.13}
that
\bq\label{5.14}
\bx \gb+\frac{1}{6}\gb R=\frac{4\pi}{3}\frac{T}{\gb}+\gs\gb\phi_{\ga}\phi^{\ga}
+2(\gs-6)\phi^{\gag}\na_{\gag}\gb+\frac{\gs}{\gb}\na^{\mu}\gb\na_{\mu}\gb
\end{equation}
This equation can be formally solved iteratively and one writes $({\bf 5N})\,\,\gb^2
=(8\pi T/R)-[(R/6)-\gs\phi_{\ga}\phi^{\ga}]^{-1}\gb\bx\gb+\cdots$.
The first and second order solutions are
\bq\label{5.15}
\gb^2_1=\frac{8\pi T}{R};\,\,\gb^2_2=\frac{8\pi T}{R}\left(1-\frac{1}{(R/6)-
\gs\phi_{\ga}\phi^{\ga}}\frac{\bx\sqrt{T}}{\sqrt{T}}+\cdots\right)
\end{equation}
To obtain the geodesic equation one uses \eqref{5.13} and assumes that
matter consists of dust with energy momentum tensor $({\bf 5O})\,\,T^{\mu\nu}
=\rho u^{\mu}u^{\nu}$ where $\rho$ (resp. $u^{\mu}$) represent matter
density (resp. matter velocity).  Putting ({\bf 5O}) into \eqref{5.13} and
multiplying by $u_{\mu}$ leads to $({\bf 5P})\,\,\hat{\na}_{\nu}(\rho u^{\nu})
-\rho(u_{\mu}\na^{\mu}\gb/\gb)=0$.  Then putting \eqref{5.13} into ({\bf 5P})
again one obtains $({\bf 5Q})\,\,u^{\nu}\hat{\na}_{\nu}u^{\mu}=(1/\gb)
(g^{\mu\nu}-u^{\mu}u^{\nu})\na_{\nu}\gb$.  Comparison now of \eqref{5.15}
and ({\bf 5Q}) with \eqref{5.10} and ({\bf 5L}) shows that one has the correct
equations for Bohmian quantum gravity provided 
\bq\label{5.16}
\gb\sim {\mf M};\,\,\frac{8\pi T}{R}\sim m^2;\,\,\frac{1}{\gs\phi_{\ga}\phi^{\ga}
-(R/6)}\sim\ga
\end{equation}
Note also that since a gauge transformation can transform a general spacetime
dependent Dirac field to a constant one and vice-versa it can be shown that the
quantum effects and the length scale of spacetime are closely related.  Thus
suppose one is in a gauge where the Dirac field is constant; then applying a 
gauge transformation one can change it to a spacetime dependent function
$({\bf 5R})\,\,\gb=\gb_0\to \gb=\gb_0exp(-\gL(x))$ via $({\bf 5S})\,\,\phi_{\mu}\to
\phi_{\mu}+\pp_{\mu}\gL$.  Thus the gauge in which the quantum mass is
constant (with zero quantum force) and the gauge in which it is spacetime
dependent are related by a scale change; i.e. $\phi_{\mu}$ in the two gauges
differs by $-\na_{\mu}(\gb/\gb_0)$.  Since $\phi_{\mu}$ is part of Weyl geometry
and $\gb$ represents the quantum mass one concludes that the quantum 
effects are geometrized.
\\[3mm]\indent
{\bf REMARK 5.2.}
The work of Quiros et al in \cite{a2,ar,bol,gz,q1,q2,q3} is very striking and we 
discussed this at some length in \cite{c1} (cf. also \cite{c5}).  One can begin in
\cite{q2} with the transformation of units theme.  One has two Brans-Dicke (BD) actions
($\hat{g}_{ab}=\gO^2(x)g_{ab},\,\,\hat{m}=\gO^{-1}(x)m$, etc.)
\bq\label{5.17}
({\bf A})\,\,S_{BD}=\int d^4x\sqrt{-g}\left[\phi R-\frac{\go}{\phi}|\na\phi|^2+16\pi L_M\right];
\end{equation}
$$({\bf B})\,\,\hat{S}_{BD}=\int d^4x\sqrt{-\hat{g}}\left[\hat{R}-\left(\go+\frac{3}{2}\right)
|\hat{\na}\psi|^2+16\pi e^{-2\psi}L_M\right]$$
($\phi\to exp(\psi),\,\,\gO^2=\phi$).  The conformal Riemann structure $\hat{g}\sim$
integrable Weyl spacetime.  According to Quiros et al the laws of spacetime should
be invariant under transformations of the group of point dependent transformations of
units (length, time, mass); this is called the BD postulate.  In addition to \eqref{5.17} one
has an Einstein frame from general relativity (GR) derived from
\bq\label{5.18}
S_{GR}=\int d^4x\sqrt{-g}(R-\ga|\na\psi|^2+16\pi L_M)
\end{equation}
(cf. \cite{bol,ma,q1}) whose conformal form (conformal GR) is
\bq\label{5.19}
\hat{S}_{GR}=\int d^4x\sqrt{-\hat{g}}e^{-\psi}\left[\hat{R}-\left(\ga-\frac{3}{2}\right)|\hat{\na}\psi|^2+16\pi e^{-\psi}L_M\right]=
\end{equation}
$$\int d^4x\sqrt{-\hat{g}}\left[\hat{\phi}\hat{R}-\left(\ga-\frac{3}{2}\right)\frac
{|\hat{\na}\hat{\phi}|^2}{\hat{\phi}}+16\pi\hat{\phi}^2L_M\right]$$
($\gO^2=exp(\psi)=\phi$ and $\hat{\phi}=exp(-\psi)=\phi^{-1}$)
and we recall $\na_cg_{ab}=0$ is transformed into $\hat{\na}_c\hat{g}_{ab}=(\pp_c\psi)
\hat{g}_{ab}$ (cf. \eqref{1.10} and note $(\hat{\na}\psi)^2=[\hat{\na}(-log(\hat{\phi})]^2=(\hat{\na}\hat{\phi})^2/\hat{\phi}^2$).  Thus Riemannian geometry is transformed into IW
geometry.  One then eventually concludes (cf. \cite{q2,q3} that the only consistent
formulation is conformal GR based on \eqref{5.18} - namely \eqref{5.19}.  This means that BD theory, which is naturally linked with Riemannian geometry, is not consistent with
unit transformations (nor is Einstein GR); however conformal GR as in \eqref{5.19}
is consistent and is naturally linked with Weyl geometry.  It appears that the 
conformal BD theory as in (5.17B) differs from conformal GR only in a parameter 
choice $\ga=\go+3$; however there is a difference in the matter Lagrangian $L_M$
and this leaves conformal GR as the only consistent formulation.
There is a great deal of discussion of these matters in \cite{a2,ar,bol,cd,cs,gz,ma,q1,q2,q3}.
In this regard note that under a conformal rescaling $({\bf 5T})\,\,\gO^2=exp(\psi)$
\eqref{5.18} becomes, following \cite{bol} ($\hat{\phi}=exp(-\psi)$)
\bq\label{5.20}
\hat{S}_{GR}=\int d^4x\sqrt{-\hat{g}}\left[\hat{\phi}\hat{R}-\frac{\ga-(3/2)}{\hat{\phi}}|\hat{\na}
\hat{\phi}|^2+16\pi\hat{\phi}^2L_M\right]
\end{equation}
as in (5.19)
where $\hat{\phi}=exp(-\psi)$ (cf. \cite{bol}) and again $\hat{\na}_c\hat{g}_{ab}=
w_c\hat{g}_{ab}$ where $w_a$ is the Weyl gauge vector ($w_a\sim\pp_a\psi\sim
-\pp_a\hat{\phi}/\hat{\phi}$). Note here that the equations of free motion in the Weyl
spacetime based on ({\bf 5T}) with $({\bf 5U})\,\,S_M=16\pi\int d^4x\sqrt{-\hat{g}}
\hat{\phi}^2L_M$ will have the form
\bq\label{5.21}
\frac{d^2x^a}{d\hat{s}^2}+\hat{\gG}^a_{mn}\frac{dx^m}{d\hat{s}}\frac{dx^n}{d\hat{s}}
-\frac{\pp_n\psi}{2}\left(\frac{dx^n}{d\hat{s}}\frac{dx^a}{d\hat{s}}-\hat{g}^{na}\right)=0
\end{equation}
which are conformal to
\bq\label{5.22}
\frac{d^2x^a}{dx^2}+\gG^a_{mn}\frac{dx^m}{ds}\frac{dx^n}{ds}=0
\end{equation}
and mass follows the rule $\hat{m}=\hat{\phi}^{-1/2}m$.  If one sets now $\hat{\phi}^{-1}=1+Q$ (following \cite{s17}) where Q is the quantum
potential (actually $\hat{\phi}^{-1}=exp(Q)$ following ({\bf 5L})), then the
last term on the left in \eqref{5.21} represents a quantum force (via $\pp_n\psi=
\pp_nlog(\hat{\phi})=\pp_nexp(Q)$).  Recall here 
$\gO^2=exp(-\psi)=\hat{\phi}^{-1}$
is the conformal factor in $\hat{g}_{ab}=\gO^2g_{ab}$ so
$({\bf 5V})\,\,\hat{g}_{ab}=\hat{\phi}^{-1}g_{ab}=exp(Q)g_{ab}=({\mf M}^2/m^2)g_{ab}$
where we have now identified $\hat{m}$ with the quantum mass ${\mf M}$ 
(cf. \cite{bol,c1,c3,c5,s35,s17,ss3}).
We recall here that there are equations similar to (5.21) in \cite{s35} with
$-(\pp_n\psi/2)\sim(\pp_n{\mf M}/{\mf M})$ so in some sense (modulo constants)
$\psi\sim log({\mf M}^{-2})\sim log(\hat{\phi})$ which implies $\hat{\phi}^{-1}\sim
{\mf M}^2\sim\gO^2$.
The arguments here now assert that a free falling test particle in Weyl spacetime
would not feel the quantum force when following the path described by \eqref{5.21}.
Thus the IW geometry implicitly contains the effects of some quantum matter.
Note that mass changes via $({\bf 5W})\,\,\hat{m}=exp[-(1/2)\psi]m\sim\hat{\phi}^{-(1/2)}m
=exp(Q/2)m={\mf M}$ as already indicated in the quantum mass formula from
\cite{s35,s17,ss3} (see here ({\bf 5K})-({\bf 5L}) where 
$(\bgs\bgs)\,\,Q=(\hbar^2/m^2c^2)(\bx\sqrt{\rho}/\sqrt{\rho})$ for some ensemble density of quantum
matter $\rho=R^2$).  It would be interesting to speculate now on a connection between
dark matter and quantum matter (see Section 6).$\hfill\bs$

\section{DARK ENERGY AND DARK MATTER}
\renewcommand{\theequation}{6.\arabic{equation}}
\setcounter{equation}{0}

Connections between dark matter and Weyl-Dirac theory go back at least to
Israelit and Rosen \cite{i1,i3} while in \cite{c18} Castro discusses dark energy
and Weyl geometry (cf. also \cite{pd}).  We have already reviewed and sketched some of the material from \cite{i1,i2,i3} in \cite{c1,c5} and a brief sketch is given in Remark 3.1
(cf. also \cite{a2,ar,bn,bol,bd,c19,cd,cs,fao,ga,gz,ha,ma,pd,pb,q1,q2,q3,z1}).
\\[3mm]\indent
{\bf REMARK 6.1.}
There is considerable material concerning Weyl dark matter arising from from
$w^{\mu}$; this is in the form of a Weyl gas consisting of massive bosons of spin one.
Another form of dark matter is constructed using the $\gb$ field (cf. also \cite{c18,ti}).
A sketch of the theory for conformally coupled dark matter appears in the last paper of
\cite{i2} from which we extract here.  Thus the field equations of the Weyl-Dirac theory
can be derived via $({\bf 6A})\,\,\gd I_D=0$ where $I_D=\int L_D\sqrt{-g}d^4x$ and 
\bq\label{6.1}
L_D=W_{\mu\nu}W^{\mu\nu}-\gb^2R+k\pp_{\mu}\gb\pp^{\mu}\gb+(k-6)(2\gb w^{\mu}
\gb_{\mu}+\gb^2w_{\mu}w^{\mu})+2\gL\gb^4
\end{equation}
Dirac took $k=6$ in which case $w_{\mu}$ can be interpreted as the vector potential
of the EM field and $W_{\mu\nu}$ becomes the field tensor; this leads to
$({\bf 6B})\,\,L_D=W_{\mu\nu}W^{\mu\nu}-\gb^2R+6\gb^{\gs}\gb_{\gs}+2\gL\gb^4$.
For cosmological purposes one now neglects the Maxwell term so that $({\bf 6C})\,\,
L_D=-\gb^2R+6\gb^{\gs}\gb_{\gs}+2\gL\gb^4$.  Putting this into ({\bf 6A}) one has the
action for a conformal space and combining with the E-H action for gravitation and
matter $({\bf 6D})\,\,L_G=\int (R+L_M)\sqrt{-g}d^4x$ leads to
\bq\label{6.2}
I=\int [R+L_M+8\pi(\gb^2R+(1/3)\gL\gb^4)]\sqrt{-g}d^4x
\end{equation}
This is the general covariant action for a scalar field $\gb(x)$ coupled conformally
to gravitation; one assumes here that $L_M$ does not depend on $\gb$ while
dark matter is obtained from $\gb$.  Now varying the metric tensor in \eqref{6.2}
gives $({\bf 6E})\,\,R_{\mu\nu}-(1/2)g_{\mu\nu}R=-8\pi(T_{\mu\nu}+\Xi_{\mu\nu})$
where $({\bf 6F})\,\,8\pi\sqrt{-g}T_{\mu\nu}=\gd(\sqrt{-g}L_M/\gd g^{\mu\nu})$ and the
dark matter comes from the energy momentum density of the $\gb$ field
\bq\label{6.3}
6\Xi_{\mu\nu}=4\gb_{\mu}\gb_{\nu}-g_{\mu\nu}\gb^{\gs}\gb_{\gs}-2\gb\na_{\nu}\gb_{\mu}
+2g_{\mu\nu}\gb\na_{\gs}\gb^{\gs}=\gb^2(R_{\mu\nu}-(1/2)g_{\mu\nu}R)
\end{equation}
Further variation in $\gb$ yields the field equations
\bq\label{6.4}
\na_{\gs}\gb^{\gs}+(1/6)\gb R=0\Rightarrow \Xi_{\gs}^{\gs}=0\Rightarrow R=8\pi T_{\gs}^
{\gs}
\end{equation}
Hence the field equation in \eqref{6.4} becomes $({\bf 6G})\,\,\na_{\gs}\gb^{\gs}
=-(4\pi /3)\gb T^{\gs}_{\gs}$ so that the behavior of the conformally coupled field
depends on the amount and state of ordinary matter.  Finally using the contracted
Bianchi identity one obtains from \eqref{6.3}
$({\bf 6H}) \na_{\nu}\Xi_{\mu}^{\nu}=0$ and from ({\bf 3O}) 
the equation $({\bf 6I})\,\,\na_{\nu}T_{\mu}^{\nu}=0$.
$\hfill\bs$
\\[3mm]\indent
{\bf REMARK 6.2.}
In order to indicate relations to Brans-Dicke (BD) theory, which arises in some
treatments of dark energy as in Remark 6.1, we review briefly
the Weyl-Dirac theory following \cite{d1}
(sketched also in \cite{c1,c5}).  One starts with $ds'=\gag ds,\,\, ds^2=g_{\mu\nu}
dx^{\mu}dx^{\nu}\to \gag^2 ds^2$ ($\gag\sim exp(\gl)$), etc. and $\sqrt{-g}$ has weight
(or power) 4.  If T is a tensor and $T\to \gag^NT$ it is called a cotensor of power N.  One
takes a connection $({\bf 6J})\,\,{}^*\gG^{\ga}_{\mu\nu}$ as in \eqref{4.5} (or
\eqref{1.7}) so that $({\bf 6K})\,\,\na^*_{\mu}S=\hat{\na}_{\mu}S-N\phi_{\mu}S$ as in 
({\bf 4H}) where $\hat{\na}_{\mu}$ is based on $\hat{\gG}^{\gl}_{\mu\nu}\sim
\ddot{\gG}^{\gl}_{\mu\nu}$.  Then in particular ($\na^*\sim{}^*\na$)
\bq\label{6.5}
\na^*_{\nu}A=\pp_{\nu}A_{\mu}-\hat{\gG}^{\gl}_{\mu\nu}A_{\gl}-N\phi_{\nu}A_{\mu}=
\end{equation}
$$=\na_{\nu}A_{\mu}-(N-1)\phi_{\nu}A_{\mu}+\phi_{\mu}A_{\nu}-g_{\mu\nu}\phi^{\ga}
A_{\ga}$$
(co-covariant derivative of $A_{\mu}$) and also $({\bf 6L})\,\,R^*=R-6\na_{\gs}\phi^{\gs}+6\phi^{\gs}\phi_{\gs}=K$
as in ({\bf 1V}).  For the vacuum action Dirac takes
\bq\label{6.6}
I=\int \left[\frac{1}{4}W_{\mu\gl}W^{\mu\gl}-\gb^2R^*+k\na_*^{\mu}\gb
\na^*_{\mu}\gb+c\gb^4\right]\sqrt{-g}d^4x
\end{equation}
which, via ({\bf 6L}) and $\na_{\mu}(\gb^2\phi^{\mu})\to 0$ leads to
\bq\label{6.7}
I=\int\left[\frac{1}{4}W_{\mu\gl}W^{\mu\gl}-\gb^2R+6\gb^{\mu}\gb_{\mu}+c\gb^4\right]
\sqrt{-g}d^4x
\end{equation}
where $\gb_{\mu}\sim\pp_{\mu}\gb$ and $k=6$.  This agrees with \eqref{2.2} (up to a factor of 1/4) when $L_M$ is omitted and $c\sim 2\gL$.  This vacuum action no longer involves
the $\phi_{\mu}$ explicitly but only via $F_{\mu\nu}$ and is therefore invariant under
transformations $\phi_{\mu}\to \phi_{\mu}+\pp_{\mu}\chi$.  The Einstein gauge involves
$\gb=1$ and yields the standard Einstein equations.  Variation of the action gives
$({\bf 6M})\,\,\gd I=\int [(1/2)P^{\mu\nu}\gd g_{\mu\nu}+Q^{\mu}\gd\phi_{\mu}
+S\gd\gb]\sqrt{-g}d^4x$ where the term $c\gb^4\sqrt{-g}$ has been dropped on the
grounds of only being important for cosmology.  One obtains then
(calculations are omitted and perfect differentials are neglected)
\bq\label{6.8}
\gd[(1/4)F_{\mu\nu}F^{\mu\nu}\sqrt{-g}]=(1/2)E^{\mu\nu}\sqrt{-g}\gd g_{\mu\nu}-
J^{\mu}\sqrt{-g}\gd\phi_{\mu};
\end{equation}
$$E^{\mu\nu}=\frac{1}{4}g^{\mu\nu}F^{\ga\gb}F_{\ga\gb}
-F^{\mu\ga}F^{\nu}_{\ga};\,\,J^{\mu}=\hat{\na}_{\nu}F^{\mu\nu}=\sqrt{-g}^{-1}\pp_{\nu}
(F^{\mu\nu}\sqrt{-g})$$
Further calculation (continuing to neglect perfect differentials) gives then
\bq\label{6.9}
Q^{\mu}=-J^{\mu};\,\,S=-2\gb R=12\hat{\na}_{\mu}\gb^{\mu};
\end{equation}
$$P^{\mu\nu}=E^{\mu\nu}+\gb^2(2R^{\mu\nu}-g^{\mu\nu}R)-4g^{\mu\nu}\gb\hat{\na}_{\rho}
\gb^{\rho}+4\gb\hat{\na}_{\nu}\gb^{\mu}+
2g^{\mu\nu}\gb^{\gs}\gb_{\gs}-8\gb^{\mu}\gb^{\nu}$$

The field equations for the vacuum are then $({\bf 6N})\,\,P^{\mu\nu}=0;\,\,Q^{\mu}=0;\,\,
S=0$ which are not all independent since $({\bf 6O})\,\,P^{\gs}_{\gs}=-2\gb^2R-
12\gb\hat{\na}_{\gs}\gb^{\gs}=\gb S;$ thus the S equation is a consequence of the P 
equations.  If one omits the EM term from the action it becomes the same as the
Brans-Dicke action with integrand $({\bf 6P})\,\,\psi R-(\go/\psi)\pp_i\psi\pp^i\psi$
where e.g. $\psi\sim -\gb^2$
(cf. \cite{bd}) except that the latter allows an arbitrary value for the
parameter $k$ leading to an additional field equation, namely
$\bx\gb^2=0$. $\hfill\bs$
\\[3mm]\indent
{\bf REMARK 6.3.}
Dark energy and quintessence are treated in \cite{ar,bp,cd,cs,c18,ce,gz,g1,z1} for example and 
often models of the form 
\bq\label{6.10}
S=\int d^4x\sqrt{-g}\left[\frac{c^2}{16\pi G}(R-2\gL)+L_{\phi}+L_M\right]
\end{equation}
are used, where $L_{\phi}$ involves a quintessence field $({\bf 6Q})\,\,L_{\phi}=
-(1/2)\pp_n\phi\pp^n\phi-V(\phi)$.  One then often goes directly to an FRW universe
with metric $({\bf 6R})\,\,ds^2=-dt^2+a^2(t)\gd_{ik}dx^idx^k$ for example.  In \cite{c19}
Castro uses a Jordan BD action with a Jordan BD field $\phi$ of Weyl weight $-1$
\bq\label{6.11}
S=-\int d^4x\sqrt{|g|}[\phi^2({}^WR)-(1/2)g^{\mu\nu}D_{\mu}\phi D_{\nu}\phi-V(\phi)]
\end{equation}
where $D_{\mu}\phi=\pp_{\mu}\phi-A_{\mu}\phi$ and ${}^WR=R-6A_{\mu}A^{\mu}
+6\na_{\mu}A^{\mu}$ (with signature $(+,-,-,-)$ - recall $K=\hat{R}-6\na_{\gl}w^{\gl}
+6w^{\gl}w_{\gl}$ with signature $(-,+ + +)$ so $K\sim-{}^WR$ via $R\to -R$ under
signature change).
We omit further discussion for the
moment.$\hfill\bs$
\\[3mm]\indent
{\bf REMARK 6.4.}
We turn now to dark matter following Israelit \cite{i1,i2} and go back to the formulas of
Section 2; thus using the action \eqref{2.1} with no $W_{\gl\mu}$ and a slight change
in parameter expression leads to ($R^{\gs}_{\gs}=R$)
\bq\label{6.12}
I=\int d^4x\sqrt{-g}[-\gb^2R_{\gs}^{\gs}-16\pi\gk^2(\gb^2w^{\gs}w_{\gs}+
\end{equation}
$$+2\gb w^{\gs}
\pp_{\gs}\gb+g^{\gl\gs}\pp_{\gl}\gb\pp_{\gs}\gb)+6g^{\gl\gs}\pp_{\gl}\gb\pp_{\gs}\gb
+2\gL\gb^4+L_M]$$
(here $\gs+72\pi\gk^2=3$ using $2\gb_{\mu}=\gb w_{\mu}$ as in ({\bf 6S})
below).
We want to keep in mind here that the Dirac field $\gb$ may be identified with a quantum
mass ${\mf M}\sim\gb$ as indicated in \eqref{5.16}.  We have also seen in Remark 5.2
that conformal Einstein gravity (i.e. conformal GR) with $\hat{\phi}=exp(-\psi)$ and
$\gO^2=exp(\psi)$ also exhibits quantum mass via $\hat{\phi}^{-1/2}=exp(Q/2)$.  Thus
$\hat{g}_{ab}=\gO^2g_{ab}$ with ${\mf M}^2/m^2=\gO^2=\hat{\phi}^{-1}$ which implies
${\mf M}^2=\gb^2=\hat{\phi}^{-1}m^2$ or
$\gb=m\hat{\phi}^{-1/2}$
and conformality would then automatically introduces a Dirac 
field in addition to a gauge field as indicated after (5.20) (see Remark 6.5).
Consequently the step to consistent units transformation properties via 
conformality leads automatically to Weyl geometry, to quantum mass,
and to some kind of Dirac-Weyl theory.  Therefore
Weyl spacetime has quantum mass in its very existence and insofar as Weyl geometry
produces also dark matter as in \cite{i1,i2} (see below) it seems to mean that there is a relation
between quantum matter and dark matter (note this is all implicit in \cite{c1,c5}).
Note also that $Q/2=-(1/2)log(\hat{\phi})=(1/2)\psi=log(\gb/m)$ so formulas in $\hat{\phi}$
go directly into formulas in Q with no intervening mass term $m$
(also $Q=2log(\gb/m)\Rightarrow exp(Q)=\gb^2/m^2$).  In particular 
(5.19)-(5.20) become ($\hat{\phi}\sim exp(-Q)\equiv \psi= Q$)
\bq\label{6.13}
\hat{S}_{GR}=\int d^4x\sqrt{-\hat{g}}\left[e^{-Q}\hat{R}-\left(\ga-\frac{3}{2}\right)e^{Q}
(\hat{\na}e^{-Q})^2+16\pi e^{-2Q}L_M\right]=
\end{equation}
$$=\int d^4x\sqrt{-\hat{g}}e^{-Q}\left[\hat{R}-\left(\ga-\frac{3}{2}\right)(\hat{\na}Q)^2+16\pi e^{-Q}L_M\right]$$
Note \eqref{5.20} is conformal GR obtained from \eqref{5.18} via a conformal rescaling
which of necessity leads to a relation between $\hat{\phi}$ and Q via connection
to an IW theory with $\gO^2={\mf M}^2/m^2=exp(Q)$; moreover via ({\bf 5W}) we
obtain ${\mf M}=\hat{m}$.
The W-D theory of Section 5 with action as in (5.3) also produces a conformal
factor ${\mf M}^2/m^2$ and a quantum mass ${\mf M}\sim\gb$ so in some sense
(6.11)
is equivalent to a Weyl-Dirac theory with action as in (5.2) and $\gb\sim {\mf M}$. 
Thus perhaps we should try to put \eqref{5.20} into a suitable Weyl-Dirac form (based on
e.g. \eqref{6.12} or \eqref{2.1}.  Recall from $(\bgs)$ (cf. Remark 7.1)
\bq\label{6.14}
-\gb^2K=-\gb^2\hat{R}=-\gb^2R+6\gb^2\na_{\gl}w^{\gl}-6\gb^2w^{\gl}w_{\gl}
\end{equation}
with $\gb^2\na_{\gl}w^{\gl}\to -2\gb w^{\gl}\pp_{\gl}\gb$ after integration, leading to
a version of \eqref{2.1}
\bq\label{6.15}
I=\int d^4x\sqrt{-g}[-\gb^2R+\gs\gb^2 w^{\gl}w_{\gl}+(\gs+6)\pp_{\gl}\gb\pp^{\gl}\gb
+2\gs \gb w^{\gl}\pp_{\gl}\gb+2\gL\gb^4]
\end{equation}
We would like to find a W-D action with $\hat{m}\sim m\hat{\phi}^{-1/2}$ which arises from
a gravitational situation, involving an equation of the form (2.1), (2.2), (3.14), (5.3), (6.7),
or (6.12).  Thus in addition to (6.15) $\sim$ (2.1) consider the essential integrands
($\gb_{\mu}=\pp_{\mu}\gb$)
\bq\label{6.16}
-\gb^2R+k\gb_{\mu}\gb^{\mu}+(k-6)(2\gb w^{\mu}\gb_{\mu}+\gb^2w_{\mu}w^{\mu}
+2\gL\gb^4\sim (5.3);
\end{equation}
$$-\gb^2R +6\gb^{\mu}\gb_{\mu}+c\gb^4\sim (2.2)\sim (6.7);$$
$$-\gb^2R -16\pi\gk^2(\gb^2w^{\gs}\gb_{\gs}+2\gb w^{\gs}\gb_{\gs}+\gb^{\gs}\gb_{\gs})
+6\gb^{\gs}\gb_{\gs}+2\gL\gb^4\sim (6.12);$$
$$\sqrt{-\hat{g}}\left[\hat{\phi}\hat{R}-\frac{[\ga-(3/2)]}{\hat{\phi}}(\hat{\na}\hat{\phi})^2
\right]\sim (5.20)$$
(note $\hat{\phi}^2\sqrt{-\hat{g}}\sim\sqrt{-g}$ via $\hat{g}_{ab}=\gO^2g_{ab}
\equiv g_{ab}=\hat{\phi}\hat{g}_{ab}$).
Note also that $({\bf 6Q})\,\,w_{\mu}=\pp_{\mu}\psi=-\hat{\phi}_{\mu}/\hat{\phi}$ and $({\bf 6R})\,\,\gb=m\hat{\phi}^{-1/2}\Rightarrow \gb_{\mu}=(\gb/2)(-\hat{\phi}_{\mu}/\hat{\phi})$.  Consequently $({\bf 6S})\,\,\gb_{\mu}=(1/2)(w_{\mu}\gb)\Rightarrow
\pp_{\mu}log(\gb)=(1/2)w_{\mu}$ and for example in (6.3) one has
\bq\label{6.17}
-\gb^2R+6\gb^{\mu}\gb_{\mu}+c\gb^4\sim -\gb^2R+3\gb\gb^{\mu}w_{\mu}+c\gb^4
\end{equation}
To clarify (6.17) we note that for $\gs=0$ in (6.15) one has an integrand (using
({\bf 6S}) - recall $\gb_{\mu}=\pp_{\mu}\gb$)
\bq\label{6.18}
-\gb^2R+6\gb_{\gl}\gb^{\gl}+2\gL\gb^4=-\gb^2R+3\gb\gb^{\mu}w_{\mu}+2\gL\gb^4
\end{equation}
and this corresponds to the original Dirac form (6.7) $\equiv$ (2.2) so (6.18) 
comes then from Israelit's (2.1) and hence (6.12).  We also observe that the 
Shojai action in \cite{s35} (from which the $\gb={\mf M}$ result emerges) has the form
\bq\label{6.19}
A=\int d^4x\sqrt{-g}\left[W^{\mu\nu}W_{\mu\nu}-\gb^2\hat{R} +(\gs+6)
\ddot{\na}_{\mu}\gb\ddot{\na}^{\mu}\gb+L_M\right]
\end{equation}
which is the same as the Rosen action (5.4) (agreeing with (2.1)).  Hence our
calculations seem to imply the following

\section{SOME SUMMARY REMARKS}
\renewcommand{\theequation}{7.\arabic{equation}}
\setcounter{equation}{0}

The introduction of unit invariants to GR leads from classical Einstein GR to an
IW geometry in conformal GR with $\hat{g}_{ab}=\gO^2g_{ab}$.  A ``conformal" mass
then arises via $\hat{m}^2/m^2\sim\hat{\phi}^{-1}$ where $\hat{\phi}=exp(-\psi)$
and the Weyl vector is $w_{\mu}=\pp_{\mu}\psi=-(\hat{\phi}_{\mu}/\hat{\phi})$ with
action (5.20).  The conformal mass has the form $\hat{m}=m\hat{\phi}^{-1/2}$ 
following \cite{bol} and
$\hat{m}^2/m^2=\hat{\phi}^{-1}=\gO^2$.  On the other hand one can construct a
Dirac-Weyl theory with action $(2.1)\equiv (6.15)\equiv (5.3)\equiv (5.4)$ which
are also equivalent to the Shojai action (6.19) from which is derived $\gO^2=
{\mf M}^2/m^2=exp(Q)$ with $Q$ a natural quantum potential as in ({\bf 5K}) and
${\mf M}\sim \gb$. 
The identification now of $\hat{m}$ with ${\mf M}$ then leads to $\gb\sim {\mf M}=\hat{m}$ with $({\bf 6Q})\,\,2\gb_{\mu}
= w_{\mu}\gb$ and to (6.13) expressing
$\hat{S}_{GR}$ in terms of the quantum potential $Q$ as in $({\bf 5K})\sim
(\bgs\bgs)$.  This may also suggest some connections of dark matter to the ensemble
density characterizing the quantum potential via the procedures of \cite{i1,i2}.
\\[3mm]\indent
{\bf REMARK 7.1.}
As indicated after (6.20) $\sqrt{-\hat{g}}\hat{\phi}\hat{R}\leftrightarrow
\sqrt{-g}R$ in (5.20) and thus (5.20) is connected to the 
Weyl-Dirac actions above.  More precisely following e.g. \cite{capo} 
we think of conformal transformations $\hat{g}_{\mu\nu}=\gO^2g_{\mu\nu}$
(so $\go^2\sim exp(-2w)$ in \cite{capo} or $\gO^2\sim -(1/2F)$).
Then 
\bq\label{7.1}
\sqrt{-\hat{g}}(F\hat{R}+(1/2)\hat{g}^{\mu\nu}\hat{\na}_{\mu}\hat{\chi}\hat{\na}_{\nu}
\hat{\chi}-\hat{V}(\hat{\chi})=
\end{equation}
$$=\sqrt{-g}[-(1/2)R+(1/2)\na_{\ga}\chi\na^{\ga}\chi-V(\chi)]$$
Note in particular that $\sqrt{-\hat{g}}\hat{R}(2\gO^2)^{-1}\leftrightarrow
-\sqrt{-g}(R/2)$ so $\gO^2\sqrt{-g}R\sim \sqrt{-\hat{g}}\hat{R}\equiv \sqrt{-g}R
\sim \hat{\phi}\sqrt{-\hat{g}}\hat{R}$ as indicated after (6.20).
In addition one has
$({\bf 7A})\,\,(1/2)\sqrt{-\hat{g}}\hat{\na}^{\nu}\hat{\chi}\hat{\na}_{\nu}\hat{\chi}\leftrightarrow (1/2)\sqrt{-g}\na_{\ga}\chi\na^{\ga}\chi$ and going to (5.1)-(5.2)
or (6.19) for example we see that covariant derivatives go into covariant derivatives
of the same form.  Moreover from $\gb^2=m^2\hat{\phi}^{-1}=m^2exp(\psi)=
m^2\gO^2$ we can write $2\pp_a\gb/\gb=\pp_a\psi$ etc. and thus $\gb$ derivatives
go into $\psi$ derivatives times $exp(\psi/2)$, which is consistent with (5.19)
for example.  Thus it appears as if the Weyl-Dirac formulation is in fact equivalent to
that of conformal GR but some details could still deserve to be checked.
$\hfill\bs$
\\[3mm]\indent
{\bf REMARK 7.2.}  Despite all the beautiful formulas involving $Q=\ga(\bx
\sqrt{\rho}/\sqrt{\rho})$ where $\rho\sim |\Psi|^2$ for $\Psi$ a Schr\"odinger
wave function there are problems in interpretation of $|\Psi|^2$ as a 
probability density.  This has been addressed by many authors including
D\"urr, Goldstein, Zanghi, et al and Nikoli\'c (see e.g. \cite{c1,c5,nik}).
$\hfill\bs$


\newpage
\begin{thebibliography}{cccc}

%
\bibitem{ad}
R. Adler, M. Bazin, and M. Schiffer, Introduction to general relativity, McGraw-Hill,
1975
%
\bibitem{an}
L. Anderson and J. Wheeler, hep-th 0305017, 0406159, and 0412229
%
\bibitem{a2} O. Arias and I. Quiros, gr-qc 0212006
%
\bibitem{ar} O. Arias, T. Gonzalez, Y. Leyva, and I. Quiros, Class.Quantum Gravity,
20 (2003), 2563-2578
%
\bibitem{as} G. Asanov, Finsler geometry, relativity, and gauge theories, Reidel,
1985
%
\bibitem{au} J. Audretsch, Phys. Rev. D, 24 (1981), 1470-1477;
27 (1983), 2872-2884
%
\bibitem{av} J. Audretsch, F. G\"ahler, and N. Straumann, Comm. Math. Phys.,
95 (1984, 41-51
%
\bibitem{aw} J. Audretsch and C. L\"ammerzahl, Class. Quant. Gravity, 5 (1988),
12885-1295
%
\bibitem{bp} N. Banerjee and D. Pavon, Class. Quantum Gravity, 18 (2001), 593-599
%
\bibitem{bn} R. Bean, S. Carroll, and M. Trodden, astro-ph 0510059
%
\bibitem{b9} G. Bertoldi, A. Faraggi, and M. Matone, hep-th 9909201
%
\bibitem{bl} M. Blagojevi\'c, Gravity and gauge symmetry, IOP Press, 2002
%
\bibitem{bol} R. Bonal, I. Quiros, and R. Cardenas, gr-qc 0010010
%
\bibitem{bd} C. Brans and R. Dicke, Phys. Rev., 124 (1961), 925-935
%
\bibitem{c19} V. Canuto,, P. Adams, S. Hsieh, and E. Tsiang, Phys. Rev. D, 16
(1977), 1643-1663
%
\bibitem{capo} S. Capozziello and M. Francaviglia, astro-ph 0706.1146
%
\bibitem{cd} R. Cardenas, T. Gonzalez, O. Martin, and I. Quiros, astro-ph 0210108
%
\bibitem{cs} R. Cardenas, T. Gonzalez, Y. Leiva, and I. Quiros, astro-ph 0206315
%
\bibitem{c1} R. Carroll, Fluctuations, information, gravity, and the quantum potential, Springer, 2006
%
\bibitem{c2} R. Carroll, gr-qc 0501045
%
\bibitem{c3} R. Carroll, math-ph 0701007
%
\bibitem{c4} R. Carroll, math-ph 0703065
%
\bibitem{c5} R. Carroll, On the quantum potential, Abramis Academic, to appear
%
\bibitem{c16} C. Castro, Found. Phys., 22 (1992), 569-615; Found. Phys. Lett.,
4 (1991), 81-99; Jour. Math. Phys., 31 (1990), 2633-2636
%
\bibitem{c17} C. Castro and J. Mahecha, Prog. Phys., 1 (2006), 38-45
%
\bibitem{c18} C. Castro, On dark energy, Weyl's geometry, ..., Preprint, 2006
%
\bibitem{ce} H. Cheng, Phys. Rev. Lett., 61 (1988), 2182-2184
%
\bibitem{d1} P. Dirac, Proc. Royal Soc. London A, 332 (1973), 403-418
%
\bibitem{dt} W. Drechsler and H. Tann, gr-qc 9802044
%
\bibitem{d2} W. Drechsler, Found. Phys., 29 (1999), 1327-1369
%
\bibitem{fao} V. Faraoni, Cosmology in scalar tensor gravity, Kluwer, 2004
%
\bibitem{fa} A. Faraggi and M. Matone, Inter. Jour. Mod. Phys. A, 15 (2000),
1869-2017
%
\bibitem{fe} A. Feoli, W. Wood, and G. Papini, gr-qc 9805035
%
\bibitem{fjm} Y. Fujii and K. Maeda, The scalar tensor theory of gravitation,
Cambridge Univ. Press,2003
%
\bibitem{fu} T. Fulton, F. Rohrlich, and L. Witten, Rev. Mod. Phys., 34 
(1962), 442-457
%
\bibitem{ga} R. Gannouji, D. Polarski, and A. Ranquet, astro-ph 0701650
%
\bibitem{gz} T. Gonzalez, G. Leon, and I. Quiros, astro-ph 0502383 and 0702227
%
\bibitem{g1} E. Guendelman and a. Kaganovich, Phys. Rev. D, 53 (1996),
70207025; 55 (1997), 5970-5980; 60 (1999), 065004-1
%
\bibitem{ha} A. Heavens, T. Kitching, and L. Verde, astro-ph 0703191
%
\bibitem{h1} F. Hehl, P. von der Heyde, G. Kerlic, and J. Nester, Rev. Mod.
Phy., 48 (1976), 393-416
%
\bibitem{h2} F. Heyl and Y. Obukhov, gr-qc 0001010, 0103020, and 0508029;
Found. Phys., 35 (2005), 2007-2025
%
\bibitem{h3} F. Heyl, Y. Itin, and Y. Obukhov, physics 0610221
%
\bibitem{i1} M. Israelit, The Weyl-Dirac theory and our universe, Nova
Science Publ., 1999
%
\bibitem{i2} M. Israelit, Found. Phys., 28 (1998), 205-228, 29 (1999)
1303-1322, 32 (2002),
295-321 and 945-961; gr-qc 9608035
%
\bibitem{i3} M. Israelit and N. Rosen,  Found. Phys., 22 (1992), 555-568; 24 (1994),
901-915; 25 (1995), 763
%
\bibitem{i4} M. Israelit, Gen. Relativity and Gravitation, 29 (1997), 1411-1424 and
1597-1614
%
\bibitem{l1} E. Lord, Tensors, relativity, and cosmology, Tata McGraw-Hill, 1976
%
\bibitem{ma} G. Magnano and L. Sokolowski, Phys. Rev. D, 50 (1994), 5039-5059
%
\bibitem{m1} O. Moritsch and M. Schweda, hep-th 9405133
%
\bibitem{nik} H. Nikoli/'c, quant-ph 0406173 and 06020254; hep-th 0610138
%
\bibitem{n1} J. Noldus, gr-qc 0508104
%
\bibitem{o1} B. O'Neill, Semi-Riemannian geometry, Academic Press - Elsevier,
1983
%
\bibitem{pd} T. Padmanabhan, astro-ph 0603114; gr-qc 0705.2533
%
\bibitem{pa} G. Papini, gr-qc 0304082
%
\bibitem{pb} P. Peebles, astro-ph 0207347
%
\bibitem{pe} V. Perlick, Class. Quant. Gravity, 8 (1991), 1369-1385
%
\bibitem{p1} N. Poplawski, gr-qc 0612193, 0701176, 0702129, and 0705.0351
%
\bibitem{q1} I. Quiros, Phys. Rev. D, 61 (2000), 124026; hep-th 0009169
%
\bibitem{q2} I. Quiros, hep-th 0009169 and 0010146; gr-qc 9904004 and 0004014
%
\bibitem{q3} I. Quiros, R. Bonal, and R. Cardenas, gr-qc 9905071 and 0007071
%
\bibitem{r1} N. Rosen, Found. Phys., 12 (1982), 213-224; 13 (1983), 363-372
%
\bibitem{sa} E. Santamato, Phys. Rev. D, 29 (1984), 216-222; 32 (1985)
2615-2621
%
\bibitem{schz} E. Scholz, astro-ph 0403446; gr-qc 0511113 
%
\bibitem{sczh} E. Scholz, astro-ph 0409635; gr-qc 0703102
%
\bibitem{s35} F. and A. Shojai, gr-qc 0306099 and 0404102
%
\bibitem{s17} F. and A. Shojai and M. Golshani, Mod. Phys. Lett. A, 13 (1998),
2725, 2915, and 2965
%
\bibitem{ss3} F. Shojai and M. Golshani, Inter. Jour. Mod. Phys. A, 13 (1998),
677-693 and 2135-2144
%
\bibitem{ti} S. Tiwari, gr-qc 0307079
%
\bibitem{w2} H. Wei and R. Cai, astro-ph 0607064
%
\bibitem{wyl} H. Weyl, Raum-Zeit-Materie, 1923
\%
\bibitem{w1} J. Wheeler, Phys. Rev. D, 41 (1990), 431-441
%
\bibitem{w2} W. Wood and G. Papini, gr-qc 9612042
%
\bibitem{z1} W. Zimdahl, gr-qc 0705.2131
%





\end{thebibliography}
\end{document}